\documentclass{egpubl}
\usepackage{eg2025}
 
\ConferencePaper        
\CGFccby

\usepackage[T1]{fontenc}
\usepackage{dfadobe}  

\usepackage{cite}  
\BibtexOrBiblatex
\electronicVersion
\PrintedOrElectronic

\ifpdf \usepackage[pdftex]{graphicx} \pdfcompresslevel=9
\else \usepackage[dvips]{graphicx} \fi

\usepackage{egweblnk} 

\usepackage{lineno}
\usepackage{amsmath} 
\usepackage{booktabs} 

\usepackage{cleveref}

\usepackage{xcolor} 

\usepackage{appendix}

\usepackage[ruled]{algorithm2e} 

\crefname{algocf}{alg.}{algs.}
\Crefname{algocf}{Algorithm}{Algorithms}

\makeatletter
\newcommand*{\rom}[1]{\expandafter\@slowromancap\romannumeral #1@}
\makeatother

\title[Time-dependent Persistent Wrinkles]%
      {Cloth Animation with Time-dependent Persistent Wrinkles}

\author[Deshan Gong \& Yin Yang \& Tianjia Shao \& He Wang]
{
    \parbox{\textwidth}{\centering Deshan Gong$^1$\orcid{0009-0002-2516-9542}
            Yin Yang$^{2}$\orcid{0000-0001-7645-5931}
            Tianjia Shao$^3$\orcid{0000-0001-5485-3752} 
            and He Wang$^4$\thanks{Corresponding author}\orcid{0000-0002-2281-5679} 
            } \\
    {\parbox{\textwidth}{\centering $^1$The University of Leeds UK,
        $^2$ The University of Utah USA,
        $^3$ Zhejiang University China, 
        $^4$ AI Centre, Computer Science, University College London UK
           }
    }
}

\begin{document}

\teaser{
 \includegraphics[width=\linewidth]{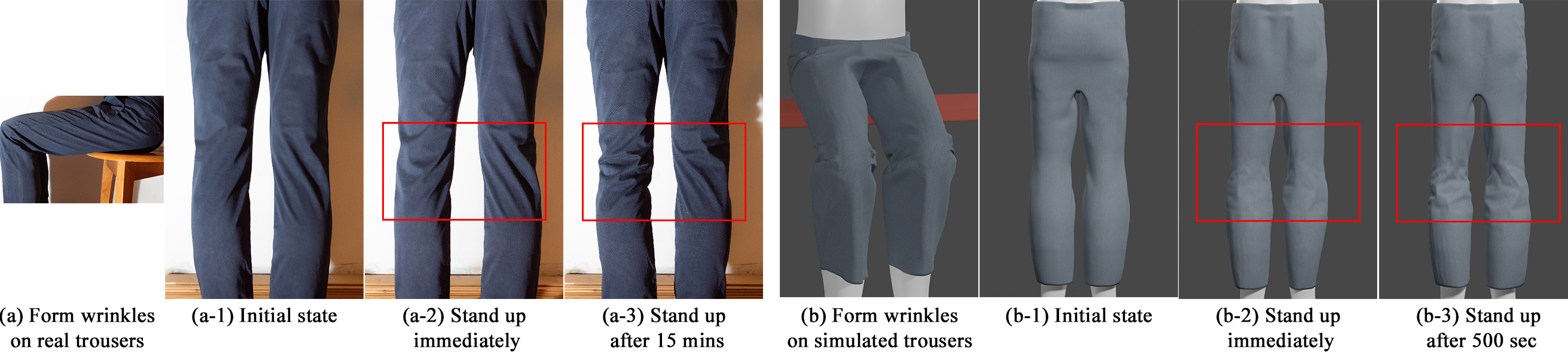}
\centering
\caption{We propose a cloth simulator that can simulate time-dependent persistent wrinkles commonly observed in real cloths. A pair of trousers start to form wrinkles around the knees after sitting down (a-2) and the wrinkles become deeper and sharper after sitting for some time (a-3). Our simulator can plausibly reproduce similar phenomena (b, b-1,2,3) with the flexibility in controlling wrinkles' time-dependent rate (i.e., keeping deformation for 500 seconds is sufficient rather than 15 minutes).}
\label{fig:teaser}
}

\maketitle
\begin{abstract}

Persistent wrinkles are often observed on crumpled garments e.g., the wrinkles around the knees after sitting for a while. Such wrinkles can be easily recovered if not deformed for long, and otherwise be persistent. Since they are vital to the visual realism of cloth animation, we aim to simulate realistic looking persistent wrinkles. To this end, we present a physics-inspired fine-grained wrinkle model. Different from existing methods, we recognize the importance of the interplay between internal friction and plasticity during wrinkle formation. Furthermore, we model their \textit{time dependence} for persistent wrinkles. Our model is capable of not only simulating realistic wrinkle patterns, but also their time-dependent changes according to how long the deformation is maintained. Through extensive experiments, we show that our model is effective in simulating realistic spatial and temporal varying wrinkles, versatile in simulating different materials, and capable of generating more fine-grained wrinkles than the state of the art. 


\end{abstract} 


\section{Introduction}

Visually realistic cloth simulation has been actively studied in computer graphics, which has been applied in animations, fashion designs, games, etc.~\cite{wang2012manipulation, stuyck2022cloth}. We investigate a novel aspect of the cloth simulation: the formation and time evolvement of the persistent wrinkles. Such wrinkles can be seen as secondary geometries to the overall cloth dynamics, but significantly affect the visual realism~\cite{bridson2005simulation}. However, their underlying physics is still under-explored.

Existing methods for wrinkle simulation can be broadly divided into three categories: rule-based, data-driven, and physics-inspired. Early attempts focus on designing rules that can map deformations into wrinkle geometries~\cite{cutler2005art}. However, to achieve visual realism, it needs a lengthy trial-and-error process, heavily relies on the users' experience and manual labor. Later, data-driven methods learn and reproduce wrinkles from data~\cite{wang2010example}, which alleviate the manual labor but cannot synthesize realistic wrinkles outside the data distribution, being particularly problematic if further simulations are needed. Alternatively, physics-inspired methods investigate the underlying physics of the wrinkle formation~\cite{rohmer2010animation}, providing a promising avenue for general wrinkle simulations.

Despite the success of physics-inspired methods, we argue that existing methods both simplify the wrinkle formation process and completely overlook its \textit{time dependence} (shown in~\Cref{fig:teaser} (a-3)). During formation, the interplay between the internal friction and the plasticity plays a key role~\cite{prevorsek1975influence}, but existing physics-inspired methods tend to attribute wrinkles to either internal friction~\cite{ngo2004nonlinear, miguel2013modeling,wong2013modelling} or plasticity~\cite{narain2013folding}. In real cloths, the former prevents the cloth from recovering after deformations by prohibiting yarns'/fibers' relative sliding~\cite{skelton196822,gong2022fine,gong2024bayesian}, while the latter causes permanent material property changes (especially in large deformations). Intuitively, wrinkles caused by the internal friction are softer than those caused by the plasticity, and they are not independent factors in the wrinkle formation. Instead, they jointly affect the formation, with their relative importance dynamically changing as the deformation magnitude changes. This dictates that both factors need to be considered simultaneously, which has been largely neglected. 

Furthermore, wrinkles are time-dependent, i.e., the longer the deformation is kept, the more obvious and persistent the wrinkles tend to be~\cite{levison1962some}. An intuitive real-world example is a crumbled shirt being pressed for a long time tends to form firm wrinkles with sharp edges. This suggests that there is a process analogous to solidification when wrinkles are formed, which needs to be modeled in the plasticity and the internal friction especially for persistent wrinkles. This time dependence has not been studied in the existing works, to the best of our knowledge.

To fill the research gap, we propose a new wrinkle model that is capable of generating visually realistic persistent wrinkles. It captures the interplay between the internal friction and the plasticity during the wrinkle formation, and explicitly models their time dependence. While the friction model is responsible for wrinkles in small deformations caused by inter-yarn sliding, the plasticity model accounts for the wrinkles caused by large deformations. Together, intensifying deformations causes wrinkles to change from friction-dominance, to a mixture of friction and plasticity, then finally to plasticity-dominance. Furthermore, both models are designed to be time-dependent, so that a longer deformation duration causes sharper, more persistent, and hard-to-recover wrinkles. Via thorough experiments, we first demonstrate that our method can simulate different wrinkles generated in small/large deformations. Next, we also simulate and compare the wrinkles generated with different durations of deformations. The results show our simulator can generate realistic time-dependent wrinkles, qualitatively compared with real garments. Further, we show our method can simulate a wide range of cloth types, from materials that are prone to hard and firm wrinkles in space and time, to materials that resist them. Through comparisons with previous work, we show that our simulator can generate visually plausible wrinkles, with the key time dependence that is commonly observed in daily life but largely missed in current research. Although our model is physics-inspired, rather than physically-based, it is the first animation technique of its kind.

Formally, our contributions include: (1) a new time-dependent friction model for cloth simulation; (2) a new time-dependent plastic model for cloth simulation; (3) the first cloth simulator, to the best of our knowledge, that can generate complex time-dependent persistent wrinkles. Code and data are shared in: \href{https://github.com/realcrane/Cloth-Animation-with-Time-dependent-Persistent-Wrinkles}{https://github.com/realcrane/Cloth-Animation-with-Time-dependent-Persistent-Wrinkles}

\section{Related Work}

Wrinkles can be separated into the \textit{dynamic wrinkles} and \textit{static/persistent wrinkles}~\cite{larboulette2004real}. The former refers to the fine geometrical details and the folds dynamically appear with cloth motions; the latter refers to the permanent deformations with which the cloth can no longer return to its original shape even under no external impact. There are various \textit{dynamic wrinkle} simulation methods which can be categorized into: rule-based methods~\cite{hadap1999animating, cutler2005art}, data-driven methods~\cite{wang2010example, lahner2018deepwrinkles}, and physics-based methods~\cite{kunii1990modeling, bridson2005simulation, wang2021gpu}. Conversely, the research in the formation of \textit{static/persistent wrinkles} is scarce. \cite{pizana2020bending} can only simulate pre-defined wrinkles rather than modeling the physics in their formations. \cite{narain2013folding,guo2018material} use plastic deformation to account for the persistent wrinkles and adopt the hardening plastic model~\cite{gingold2004discrete}. \cite{kim2011persistent} simulates permanent wrinkles by changing the rest shape and material stiffness parameters with deformations. However, they cannot simulate time-dependent wrinkles observed in real cloths \cite{levison1962some} and ignore another important factor for persistent wrinkles: internal friction~\cite{chapman197227,brenner1964mechanical,prevorsek1975influence,chapman1975importance}. \cite{miguel2013modeling} re-parameterizes Dahl's friction model to make it more suitable for cloth simulation and shows that the internal friction can also cause wrinkles. \cite{wong2013modelling} introduces a new model to simulate cloth wrinkles caused by the internal friction and fit cloth bending hysteresis behaviors: the energy loss in clothes' load-deformation processes. In yarn-level cloth simulators~\cite{cirio2014yarn}, the internal friction force is modeled as the sliding friction force between contacting yarns which prevents cloths from unraveling, and the shearing friction force which leads to shearing wrinkles. However, none of these internal friction models is time-dependent. In addition, as discovered by \cite{prevorsek1975influence}, persistent wrinkles are collectively caused by: (1) frictional fabric bending, (2) friction yarn bending, and (3) permanent bending of filaments (can be modeled by plastic deformations). However, in graphics, simulating persistent wrinkles by combining internal friction and plasticity has been rarely explored so far.

\section{Methodology}

We discretize a cloth uniformly into a triangle mesh with even face areas. The cloth state is represented by the vertex position and velocity. Specially, given a cloth mesh with $v$ vertices, we denote its state at time $t$ by $\mathcal{S}_t = \{\mathbf{x}_t, \dot{\mathbf{x}}_t\}$, where $\mathbf{x} \in \mathrm{R}^{3v}$ and $\dot{\mathbf{x}} \in \mathrm{R}^{3v}$ denote the nodal position and velocity vector respectively. Given the initial state $\mathcal{S}_0$,  the cloth motion is governed by Newton's second law, $\mathbf{f} = \mathbf{M}\ddot{\mathbf{x}}$, where $\mathbf{M} \in \mathbf{R}^{3v \times 3v}$ is the lumped mass matrix~\cite{logan2022first} and $\mathbf{f} \in \mathrm{R}^{3v}$ is the net force vector: the combined force of the internal and the external forces at vertices. To solve $\mathbf{f} = \mathbf{M}\ddot{\mathbf{x}}$, we employ implicit Euler for stability under large time steps~\cite{baraff1998large} where the the governing equation is:
\begin{equation}
\label{eq:governingEq}
    \left(
    \mathbf{M} - h^2 \frac{\partial \mathbf{f}}{\partial \mathbf{x}}  
    - h \frac{\partial \mathbf{f}}{\partial \dot{\mathbf{x}}} 
    \right)
    \Delta \dot{\mathbf{x}}_{t}
    = h \left(
    \mathbf{f}_{t-1} + h \frac{\partial \mathbf{f}}{\partial \mathbf{x}}\dot{\mathbf{x}}_{t-1}
    \right)
\end{equation}
in which $h$ is the time step size. It can be solved by an iterative solver, e.g., Conjugate Gradient \cite{shewchuk1994introduction}.

The internal force is defined by $\sigma = k \varepsilon$ with $\varepsilon$ being the strain, $\sigma$ being the stress, and $k$ being the material parameter. This model governs the out-of-plane bending and in-plane stretching forces which tend to keep the cloth in its rest state. For bending, we use an elastic model and adopt the mean-curvature~\cite{grinspun2003discrete} to define the bending strain: $\varepsilon_b = 3 \frac{\theta - \bar{\theta}}{\bar{\mathrm{H}}}$ where $\theta$ is the Dihedral angle between two adjacent triangles and $\bar{\theta}$ is the rest Dihedral angle. $\bar{\mathrm{H}}$ is the average height of the two triangles. Therefore, the elastic bending stress is $\sigma_b = K_b \varepsilon_b$ where $K_b$ is the bending stiffness. The nodal bending force is:
\begin{gather}
    W_b = A_b \int^{\varepsilon_b}_{0} \sigma_b \text{d} \varepsilon_b = A_b \int^{\varepsilon_b}_{0} K_b \varepsilon_b \text{d} \varepsilon_b = \frac{1}{2} A_b \sigma_b \varepsilon_b \label{eq:bending_force} \\
    \mathbf{f}_b = - \frac{\partial W_b}{\partial \mathbf{x}_b} = - A_b \sigma_b \frac{\partial \varepsilon_b}{\partial \mathbf{x}_b} = - 3 \frac{A_b \sigma_b}{\bar{\mathrm{H}}} \frac{\partial \theta}{\partial \mathbf{x}_b}
\end{gather}
where $W_b$ is the bending energy and $A_b = \frac{1}{3}l\bar{\mathrm{H}}$ where $l$ is the rest length of the common edge shared by the two triangles. $\mathbf{x}_b \in \mathcal{R}^{12}$ is the vertex position vector of the two triangles. Please refer to \cite{tamstorf2013discrete} for the derivatives and Jacobians of $\theta$. 

For stretching, we adopt hyperelastic Saint-Venant-Kirchhoff (StVK) constitutive model~\cite{volino2009simple} and the Green-Lagrange strain tensor to define a triangle's deformation. In Voigt Notation, the strain can be denoted by the vector $\boldsymbol{\varepsilon}_s = (\varepsilon_{uu}, \varepsilon_{vv}, \varepsilon_{uv})$. Cloths, e.g., woven fabrics, usually exhibit distinctive stretching mechanical properties in the warp and weft directions, so they are usually modeled as orthotropic materials~\cite{boisse2001analyses}. Therefore,  $\varepsilon_{uu}$, $\varepsilon_{vv}$, and $\varepsilon_{uv}$ represent the tensile strains along the warp, weft, and diagonal direction (shearing strain), respectively. The stretching constitutive equation is:
\begin{equation}
    \boldsymbol{\sigma}_s = 
    \begin{bmatrix}
        \sigma_{uu} \\ \sigma_{vv} \\ \sigma_{uv}
    \end{bmatrix}
    = \begin{bmatrix}
        k_{11} & k_{12} & 0 \\
        k_{12} & k_{22} & 0 \\
        0 & 0 & k_{33} \\
    \end{bmatrix}
    \begin{bmatrix}
        \varepsilon_{uu} \\ \varepsilon_{vv} \\ \varepsilon_{uv}
    \end{bmatrix}
    = \mathbf{K}_s \boldsymbol{\varepsilon}_s
\end{equation}
where $\mathbf{K}_s$ is the stretching stiffness matrix in which $k_{11}$, $k_{22}$,  $k_{33}$, and $k_{12}$ are the warp/weft/shear stretching stiffness and Poisson's ratio~\cite{wang2011data}. The nodal stretching force is the partial derivative of the stretching energy w.r.t. each vertex's position:
\begin{gather}
    W_s = A_s \int^{\boldsymbol{\varepsilon}_s}_0 \boldsymbol{\sigma}_s \text{d} \boldsymbol{\varepsilon}_s
    = A_s \int^{\boldsymbol{\varepsilon}_s}_0 \mathbf{K}_s \boldsymbol{\varepsilon}_s \text{d} \boldsymbol{\varepsilon}_s
    = \frac{1}{2} A_s \boldsymbol{\sigma}_s \boldsymbol{\varepsilon}_s \\
    \mathbf{f}_s = - \frac{\partial W_s}{\partial \mathbf{x}_s} 
    = -A_s  \boldsymbol{\sigma}_s \frac{\partial \boldsymbol{\varepsilon}_s}{\partial \mathbf{x}_s}
    \label{eq:stretching_force}
\end{gather}
where $W_s$ is the stretching energy, $A_s$ is the rest area of a triangle and $\mathbf{x}_s \in \mathcal{R}^{9}$ is the position vector of the triangle's vertices. Please refer to~\cite{volino2009simple} for $\frac{\partial \boldsymbol{\varepsilon}_s}{\partial \mathbf{x}_s}$ and the force Jacobians, and and the supplementary material (SM) for the other forces. We adopt~\cite{bridson2002robust} to handle collisions, and simulate the interactions and contact forces between clothes and the environment, e.g., bodies and obstacles.

\subsection{Wrinkles}

\label{sec:wrinkles}

Wrinkle formation is mainly dictated by the bending deformation \cite{wong2013modelling,wong2014improvements,fan2009engineering}. In an undeformed cloth, all the rest Dihedral angles are $\bar{\theta} = \pi$. As shown in \Cref{eq:bending_force}, the bending force always tends to keep a Dihedral angle at its rest angle $\bar{\theta}$. However, when wrinkles are formed, there are two causes: (1) another force preventing the cloth from recovering to the rest angle $\bar{\theta}$; or/and (2) the $\bar{\theta}$ being changed after deformation so that the cloth cannot return to the initial rest angle without external forces. These two causes correspond to the internal friction and the plasticity respectively.

The wrinkles that are mainly caused by the internal friction are usually soft and easy to recover (e.g., after stretching), while the ones that are mainly caused by the plasticity are often firm, persistent and even unrecoverable. For simplicity, we refer to the former as \textit{friction wrinkles} and the latter as \textit{plastic wrinkles}. 

\begin{figure}[tb]
    \centering
    \includegraphics[width=\linewidth]{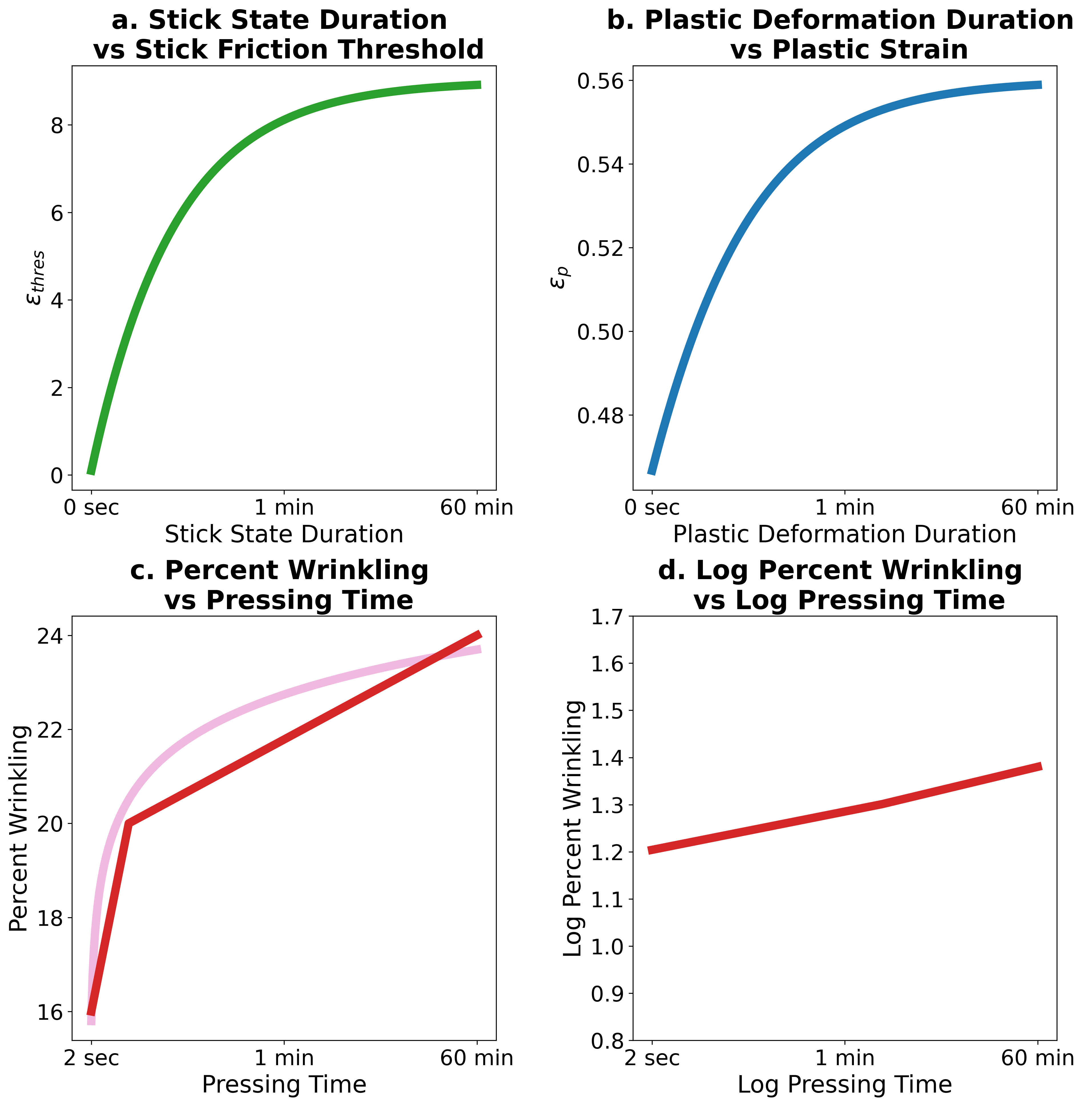}
    \caption{By using the exponential function to model friction dwell (\Cref{eq:slip-stick} and figure \textbf{a}) and time-dependent plastic hardening effects (\Cref{eq:Kh} and figure \textbf{b}), our cloth simulator can plausibly reproduce the relationship between cloth (Log) persistent wrinkles recovery percentage and (Log) pressing time (deformation duration) measured on real cloths (red curves in figure \textbf{c} and \textbf{d} from~\cite{levison1962some}). The pink curve in figure \textbf{c} approximates the measurement in \cite{levison1962some} (red curve), and its pattern is similar to the simulated curves in figure \textbf{a} and \textbf{b}.}
    \label{fig:wrinkle-time}
\end{figure}
 
\subsection{Internal Friction Model} 

\label{sec:friction}

\textbf{Shifting Anchor Friction} Friction can prevent the relative sliding between yarns/fibers. When the relative sliding is needed for the internal bending force to recover the cloth back to its rest state, yarn/fiber deformation happens hence wrinkles~\cite{lin2012finite}. We model this friction by the variation of the bending strain: $\sigma_{friction} = K_{friction} \Delta \varepsilon_b$ where $K_{friction}$ is the friction coefficient. $\Delta \varepsilon_b$ is the bending strain variation that measures the difference between the current bending strain and an anchor bending strain $\bar{\varepsilon}_b$, i.e., $\Delta \varepsilon_b = \varepsilon_{b} - \bar{\varepsilon}_b$. Note although $\sigma_{friction}$ is proportional to $\Delta \varepsilon_b$, it cannot grow unbounded as $\bar{\varepsilon}_b$ is not fixed. This is because when the bending becomes larger, the internal friction can merely counter-balance the bending force up to a threshold before sliding happens, as observed in stick-slip friction~\cite{al2004novel}. Therefore, we update $\bar{\varepsilon}_b$ when $\Delta \varepsilon_b$ is greater than a predefined stick-slip transition threshold, $\varepsilon_{thres}$:
\begin{equation}
    \label{eq:update_anchor}
    \bar{\varepsilon}_b = 
    \begin{cases}
        \bar{\varepsilon}_b &\mbox{, if } | \varepsilon_{b} - \bar{\varepsilon}_b | < \varepsilon_{thres} \\
        \bar{\varepsilon}_b + (\varepsilon_b - \varepsilon_{thres}) &\mbox{, otherwise}
    \end{cases}
\end{equation}
In the case of stick friction ($\bar{\varepsilon}_b$ not updated), the friction keeps the cloth around the anchor bending strain; otherwise the slide friction occurs ($\bar{\varepsilon}_b$ updated), the friction prevents relative motions and the anchor bending strain will move toward the current strain until $| \varepsilon_{b} - \bar{\varepsilon}_b | < \varepsilon_{thres}$. Then, it reverts back to the stick friction and keeps the cloth around the new anchor bending strain. Overall, this allows the internal friction to generate small and large wrinkles, while allowing these wrinkles to recover under external forces, e.g., flatten the cloth by stretching.

\noindent \textbf{Time dependence} One common but under-explored phenomenon in cloth is the dwell effect: the longer the stick friction state is maintained, the more difficult it is for slide friction to happen~\cite{al2004novel}. To this end, we propose a new friction component to make wrinkles time-dependent. As the duration of a friction wrinkle increases, the wrinkle becomes more difficult to recover. As $\varepsilon_{thres}$ delimits the stick-slip transition, our friction model varies this threshold with time to simulate this time-dependent behavior. Inspired by the measurements in~\cite{levison1962some} (shown in~\Cref{fig:wrinkle-time}), we use an exponential function to simulate dwell effect.
\begin{gather}
    t_{stick} =
    \begin{cases}
        t_{stick} + h, & \text{if } | \varepsilon_{b} - \bar{\varepsilon}_b | < \varepsilon_{thres} \\
        0, & \text{otherwise}
    \end{cases} \nonumber\\
    \text{where }\varepsilon_{thres} = \varepsilon_{inf} - (\varepsilon_{inf} - \varepsilon_{0}) \times e^{-\frac{t_{stick}}{\tau_f}}
    \label{eq:slip-stick}
\end{gather}
where we introduce a variable $t_{stick}$ for the duration of the stick state such that $\varepsilon_{thres}$ can vary within the interval $[\varepsilon_{0},\varepsilon_{inf}]$. Concretely, in every step, $t_{stick}$ increases by the time step size $h$ if slip friction does not occur ($| \varepsilon_{b} - \bar{\varepsilon}_b | < \varepsilon_{thres}$). In turn, this will increase $\varepsilon_{thres}$ toward $\varepsilon_{inf}$, and consequently the slip friction becomes more difficult to happen. The exponential relation is inspired by the measurements in \Cref{fig:wrinkle-time}. On the other hand, if slip friction occurs, $t_{stick}$ is reset to zero and $\varepsilon_{thres}$ is then reduced to $\varepsilon_{0}$. As a result, the slip friction resets the dwell effect. In addition, the parameter $\tau_f$ allows users to control the increase rate of $\varepsilon_{thres}$ with $t_{stick}$.

\subsection{Plastic Model}

\label{sec:plasticity}

We model plastic wrinkles through the change of the rest Dihedral angles. When a large deformation is maintained for long duration, the rest Dihedral angle changes hence the change of the cloth rest state, so that the internal bending stress does not attempt to recover the cloth to the previous rest state. To this end, we employ an elastoplastic constitutive law so the bending strain can be decoupled to an elastic part and a plastic part~\cite{o2002graphical}: $\varepsilon_b  = \varepsilon_e + \varepsilon_p$ and the bending stress is proportional to the elastic part $\sigma = K_b \varepsilon_e = K_b (\varepsilon_b - \varepsilon_p)$. Given a flat wrinkle-free cloth with no plastic strain ($\varepsilon_p = 0$), the bending stress tends to keep it wrinkle-free. Otherwise, its rest shape is changed to $\varepsilon_b - \varepsilon_p = 0$. In this case, the bending stress tends to keep the cloth in its new rest shape which is no longer flat such that plastic wrinkles appear. 

Since plasticity does not occur until the deformation reaches a threshold, referred to as the yield strain (denoted by $\varepsilon_{Y}$), a simple model is the perfect plastic model, where all of the elastic strain $\varepsilon_e$ exceeding $\varepsilon_{Y}$ is treated as the plastic strain~\cite{chaves2013notes}:
\begin{equation}
    \varepsilon_p = 
    \begin{cases}
        \varepsilon_e - \varepsilon_Y, & \mbox{if } \varepsilon_e > \varepsilon_Y \\
        \varepsilon_p, & \mbox{otherwise}
    \end{cases}  
\end{equation}
However, this simplified model does not capture the time dependence observed on real cloths~\cite{marchesi2012new}. In reality, there is a hardening process depending on time \cite{benusiglio2012anatomy}. To this end, we let the plastic deformation gradually increase with time so that plastic wrinkles can become sharper when a plastic deformation is kept for a long period. Our new time-dependent hardening plastic model alters the yield strain $\varepsilon_Y$ with time, so that the plastic strain $\varepsilon_p = \varepsilon_e - \varepsilon_Y$ becomes time-dependent. Further, we empirically observe that this dependence on time is related to the material, i.e., some materials quickly form plastic wrinkles while others take longer time. To our best knowledge, there is no generally accepted parametric form to describe this relation for various materials. But we do notice there is an overall decelerated hardening along with time across different materials~\cite{benusiglio2012anatomy}, and an exponential function can fit the measurements in~\cite{levison1962some} (\Cref{fig:wrinkle-time}). This inspires us to propose the following model. When plasticity occurs, i.e., $\varepsilon_e > \varepsilon_Y$, it changes according to:
\begin{gather}
    K_h = K_{h0} (1 - g(1-e^{(-t_{plastic}/ \tau_p)})) \label{eq:Kh} \\
    \varepsilon_{hp} \leftarrow \varepsilon_{hp} + \frac{K_b}{K_b + K_h}(|\varepsilon_e| - \varepsilon_Y) \label{eq:varhp} \\
    \varepsilon_{p} \leftarrow \varepsilon_{p} + sign(\varepsilon_e)\frac{K_b}{K_b + K_h}(|\varepsilon_e| - \varepsilon_Y) \label{eq:varp}\\
    \varepsilon_Y = \varepsilon_{Y0} + \varepsilon_{hp} \frac{K_h}{K_b} \label{eq:varY}
\end{gather}
where $K_h$ and $K_{h0}$ are the hardening parameter and initial hardening parameter respectively. $g \in (0,1)$ controls the lower bound of the hardening parameters and $\tau_p$ decides $K_h$ variation rate. The variable $t_{plastic}$ times how long the plastic deformation has been kept. In every simulation step, it is updated according to
\begin{gather}
\label{eq:tplastic}
    t_{plastic} = 
    \begin{cases}
        t_{plastic} + h &,\text{if } sign(\varepsilon_e) = sign(\varepsilon_p) \mbox{ and } \varepsilon_e > \varepsilon_Y\\
        0 &, \text{otherwise}
    \end{cases}
\end{gather}
Therefore, $t_{plastic}$ is increased by $h$ in every step if the plastic deformation flows to the overall deformation, or otherwise reduced to zero. $\varepsilon_{hp}$ denotes hardening plastic strain which is used for accumulating the plastic deformation and controlling the plastic hardening effect. Also, $\varepsilon_{p}$ denotes the actual plastic deformation, i.e., $\varepsilon_p$ in $\varepsilon_b  = \varepsilon_e + \varepsilon_p$. To simulate plastic hardening, the yield strain $\varepsilon_Y$ is affected by the plastic deformation and it can increase from a given initial yield strain $\varepsilon_{Y0}$. 

To further understand how hardening is simulated by Eq.\ref{eq:Kh}-\ref{eq:tplastic}, we describe a scenario where a plastic deformation first appears and then develops with time. When a plastic deformation first appears, $t_{plastic}$ is increased which decreases $K_h$ (\Cref{eq:Kh}). This hardens the cloth as $\varepsilon_{hp}$ is increased (\Cref{eq:varhp}), and so does the yield strain $\varepsilon_Y$ eventually (\Cref{eq:varY}). Moreover, only part of the bending strain that exceeds the yield strain is treated as plastic strain. If the current deformation is kept, $t_{plastic}$ will gradually increases and $K_h$ reduces from $K_{h0}$ toward $K_{h0}(1-g)$. Consequently, $\varepsilon_{hp}$ and $\varepsilon_Y$ further increase, and cloth is hardened. Meanwhile, the plastic strain $\varepsilon_p$ also gradually accounts more for the over all strain/stress so that the elastic strain/stress $\varepsilon_e = \varepsilon - \varepsilon_p$ reduces. This procedure is repeated as the simulation runs until $\varepsilon_e \leq \varepsilon_Y$. Overall, the longer the deformation is maintained, the larger the plastic strain is and also the more obvious the plastic wrinkles are. Refers to the SM for the algorithms, forces and Jacobian of our friction and plastic models.

\subsection{Tensile Internal Friction and Plasticity} 

Internal friction and plastic deformation also exist in cloth in-plane tensile deformations. The internal friction prevents a stretched cloth from returning to its rest length. Overly stretching a cloth can cause plastic tensile deformations so a new rest state (longer than the original one) is established and the cloth can only recover to its new rest length. Our friction and plastic model can also be applied to model cloth tensile internal friction and plasticity. Similarly to the elastic tensile model, the tensile internal friction is also orthotropic. Therefore, the tensile friction stress of a mesh triangle is:
\begin{equation}
    \boldsymbol{\sigma}_{fri} = 
    \begin{bmatrix}
        \sigma_{fri-uu} \\
        \sigma_{fri-vv} \\
        \sigma_{fri-uv}
    \end{bmatrix} = 
        \begin{bmatrix}
        K_{11} & 0 & 0\\
        0 & K_{22} & 0 \\
        0 & 0 & K_{33}
    \end{bmatrix}
    \begin{bmatrix}
        \Delta \varepsilon_{fri-uu} \\
        \Delta \varepsilon_{fri-vv} \\
        \Delta \varepsilon_{fri-uv} 
    \end{bmatrix}
\end{equation}
where $K_{11}$, $K_{22}$, and $K_{33}$ are the internal friction stiffness along the warp, the weft, and the diagonal direction respectively. The tensile internal friction strains/stresses along these directions are independent and their internal friction stiffness are updated independently in the same way as introduced in the main paper. Similarly, the tensile plastic strains in the three directions are independent as well:
\begin{equation}
    \begin{bmatrix}
        \varepsilon_{uu} \\
        \varepsilon_{vv} \\
        \varepsilon_{uv}
    \end{bmatrix} =
        \begin{bmatrix}
        \varepsilon_{e\_uu} \\
        \varepsilon_{e\_vv} \\
        \varepsilon_{e\_uv}
    \end{bmatrix} +
    \begin{bmatrix}
        \varepsilon_{p\_uu} \\
        \varepsilon_{p\_vv} \\
        \varepsilon_{p\_uv}
    \end{bmatrix}
\end{equation}
and their plastic hardening effects are also updated independently.

\section{Experiments}

Both $t_{stick}$ and $t_{plastic}$ are initialized to $0s$ in all experiments. SM provides the cloth physical parameters, implementation, additional experiments, and experimental details. Please watch the video to view the simulations. 

\subsection{Evaluation on Specimen}

\begin{figure}[htb]
    \centering
    \includegraphics[width=0.48\textwidth]{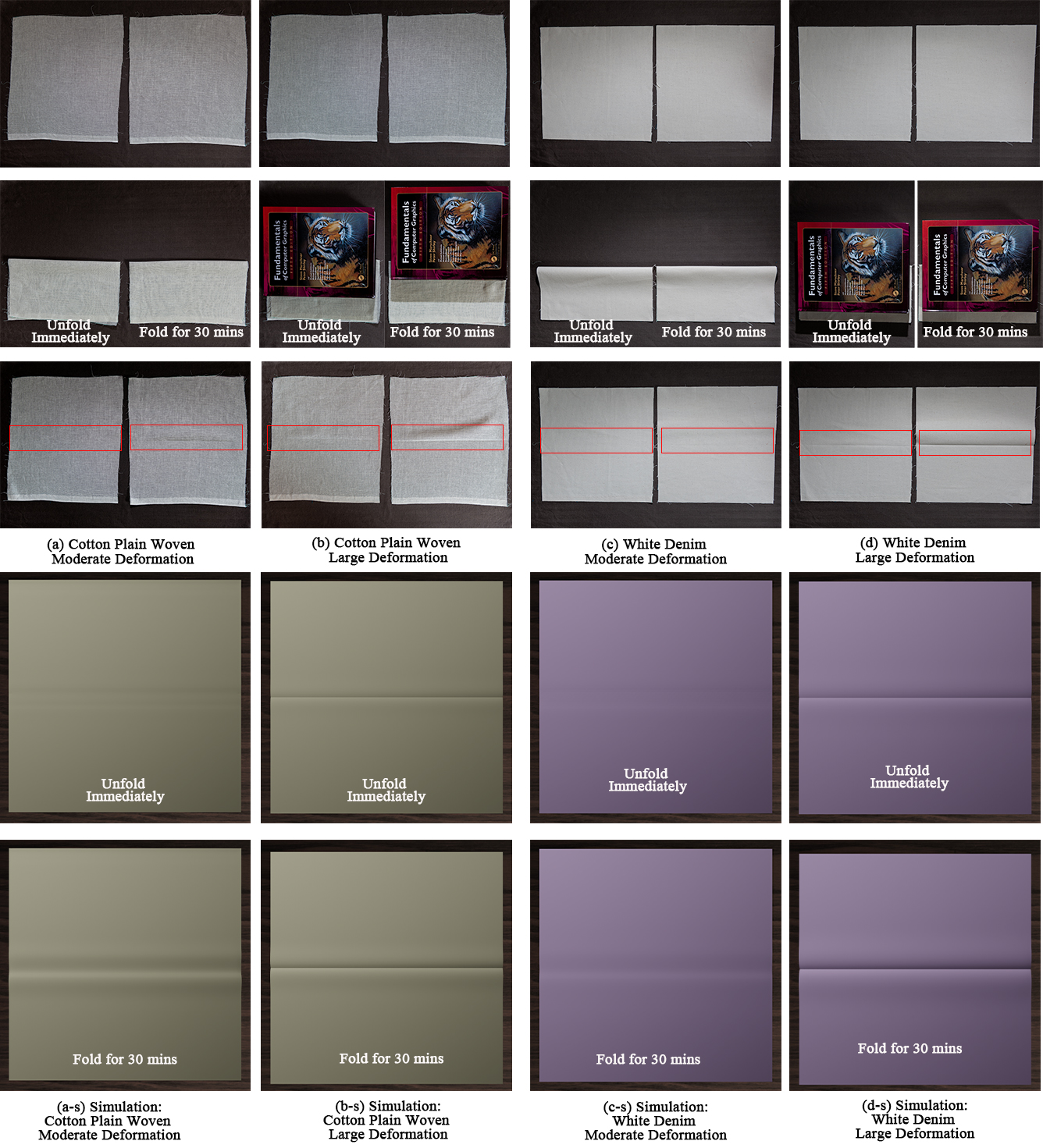}
    \caption{Our simulator can realistically simulate time-dependent wrinkles (a-s, b-s, c-s, d-s) observed in the real clothes (a, b, c, d) in different deformations. In both real and simulated clothes, folding or compressing clothes for a long duration (30 mins) makes the wrinkles sharper. The wrinkles formed on Denim by compressing (d) is sharper than those on cotton clothes (b, d), and our simulator can reproduce these observations (b-s, d-s). Bigger figures in SM.}
    \label{fig:eval_sample}
\end{figure}
\textbf{Simulate Real Materials} Our simulator can plausibly simulate the time-dependent cloth wrinkles observed in the real world. Plain-woven cotton, a.k.a. Muslin, is soft, lightweight, and breathable. It is likely to accumulate wrinkles especially in large and long-lasting deformations. As shown in~\Cref{fig:eval_sample} (a), although a short moderate folding does not cause an observe wrinkle, keeping the deformation for 30 minutes makes the wrinkle far more obvious. Moreover, extremely deforming the cloth for short time results in an obvious wrinkle and increasing the compressing time (for 30 minutes) makes the formed wrinkle even sharper (\Cref{fig:eval_sample} (b)). Thanks to our time-dependent friction model and plastic model, our simulator can simulate similar time-dependent wrinkles in either the moderate deformation~\Cref{fig:eval_sample} (a-s) or the large deformation~\Cref{fig:eval_sample} (b-s). Denim, i.e., twill-woven cotton fabric, which is thick, solid and durable, and is usually used to make jeans. Due to its large bending stiffness, it is not folded too much in the moderate deformation and therefore does not form obvious persistent wrinkles (\Cref{fig:eval_sample} (e)). Our simulation results conform to the observations (\Cref{fig:eval_sample} (e-s, f-s)). Conversely, as long as the deformation exceeds the yield strain and causes plastic deformation, the formed wrinkles is very firm and increasing the deformation duration tends to make the wrinkles more solid (\Cref{fig:eval_sample} (f)). Finally, different from plain-woven cotton, polyester (commonly used to make sport wears) is soft, light, and strongly resists to persistent deformations and even the large deformation can only cause weak wrinkle. As shown in the Figure in SM, our simulator can also realistically reproduce them. In SM, we provide the simulation parameters and also compare the side view of the real clothes' photos with our simulations. Additionally, our friction model and plastic model allow users to easily tweak the friction dwell and plastic hardening rate by tuning the $\tau_f$ and $\tau_p$ respectively. In our experiments, we set $\tau_f=30\;s$ and $\tau_p=30\;s$ such that wrinkles' dwell effect and hardening effect would become obvious as long as the duration of deformation over 30 seconds. In this way, we can simulate wrinkles' time dependence by only keeping the deformation for 500 seconds instead of an overly long time.

\begin{figure}
    \centering
    \includegraphics[width=0.48\textwidth]{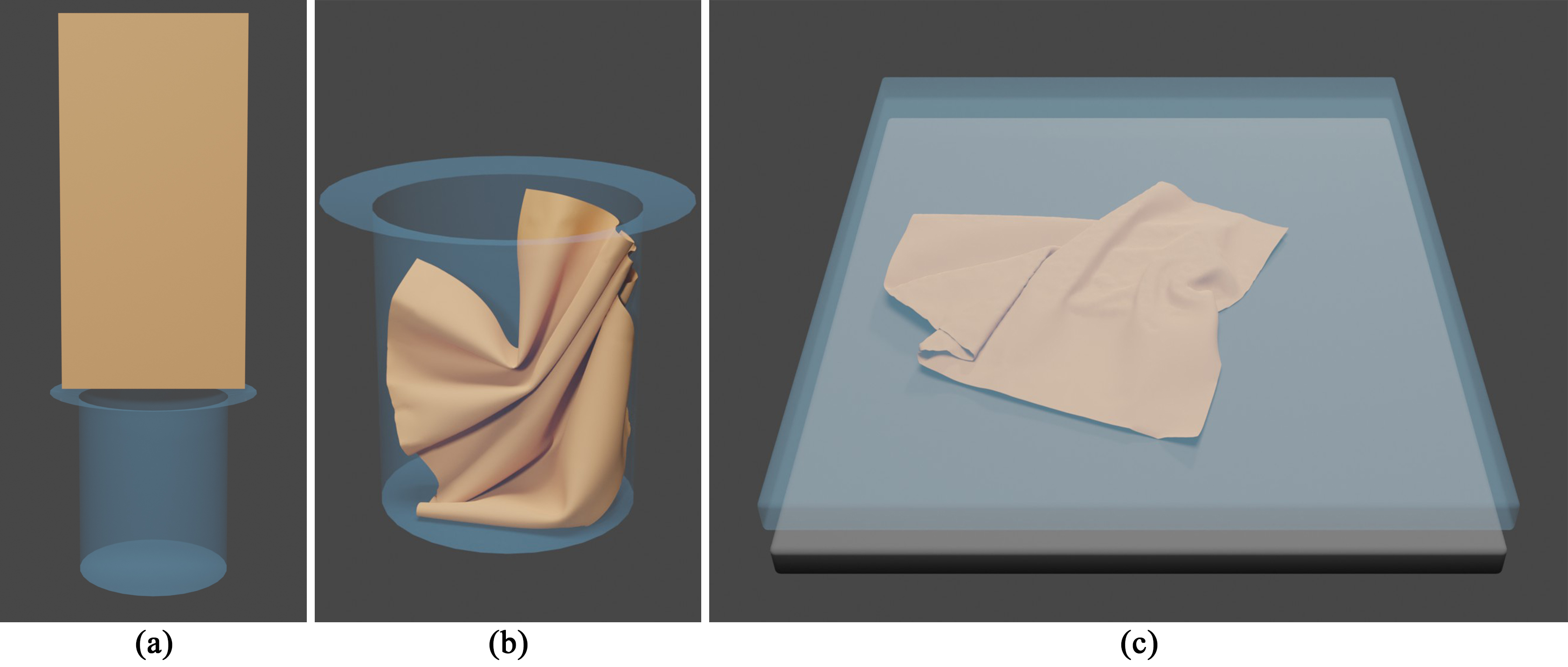}
    \caption{(a) An originally wrinkleless rectangle cloth falls into a cylindrical container due to its self-weight; (b) The cloth is folded moderately due to collision; (c) To cause extreme deformations, we compress the cloth after it falls on the ground.}
    \label{fig:motion_fall}
\end{figure}
\begin{figure}
    \centering
    \includegraphics[width=0.48\textwidth]{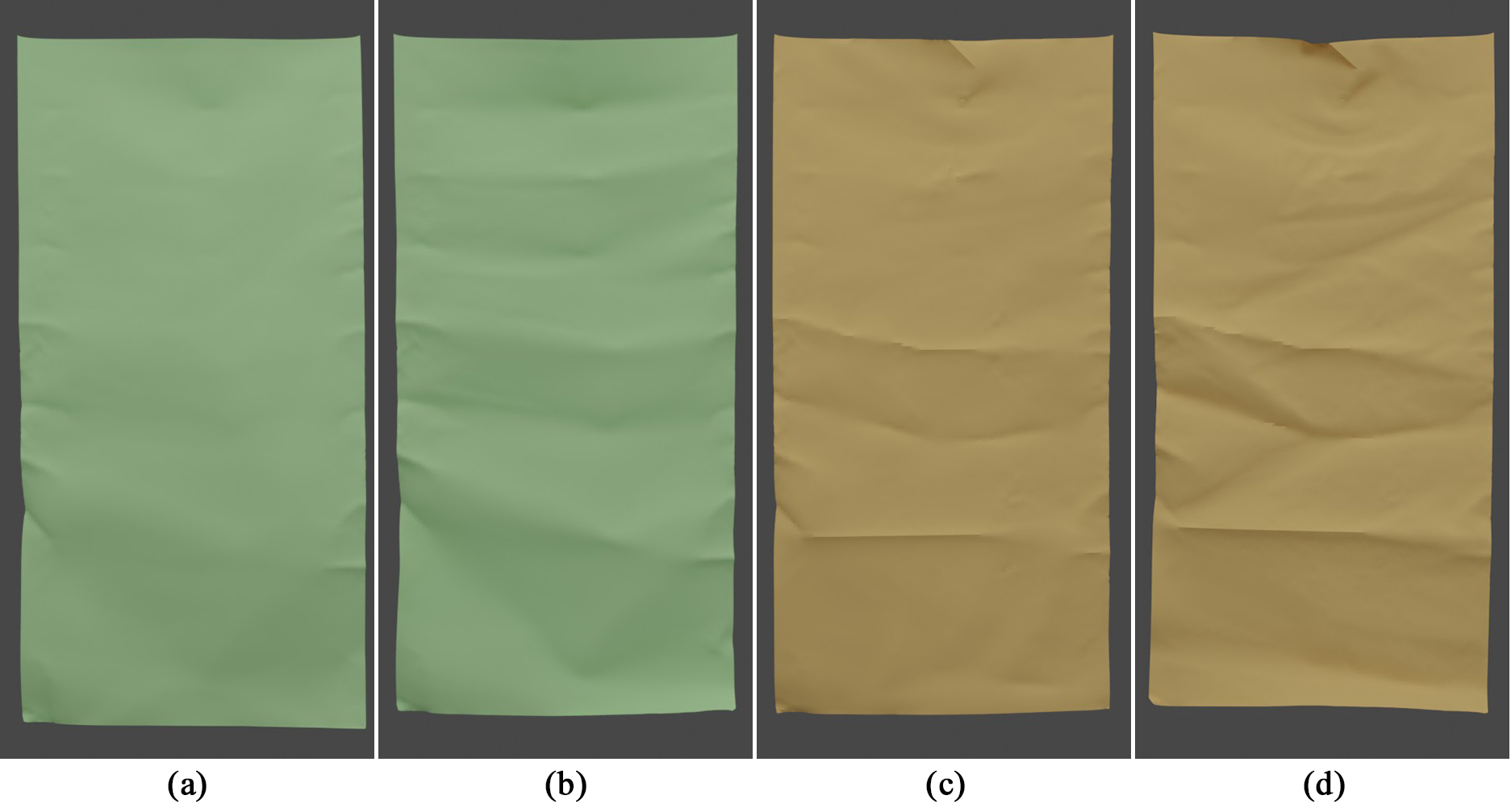}
    \caption{The wrinkles on the cloth after lifting it. (a) Immediately lift the cloth after folding moderately; (b) Lift the cloth after keeping the moderate deformation for 500s; (c) Immediately lift the cloth after being compressed by the heavy weight; (d) Lift the cloth after compressing it for 500s.}
    \label{fig:cloth_wrinkles}
\end{figure}
\begin{figure}
    \centering
    \includegraphics[width=0.48\textwidth]{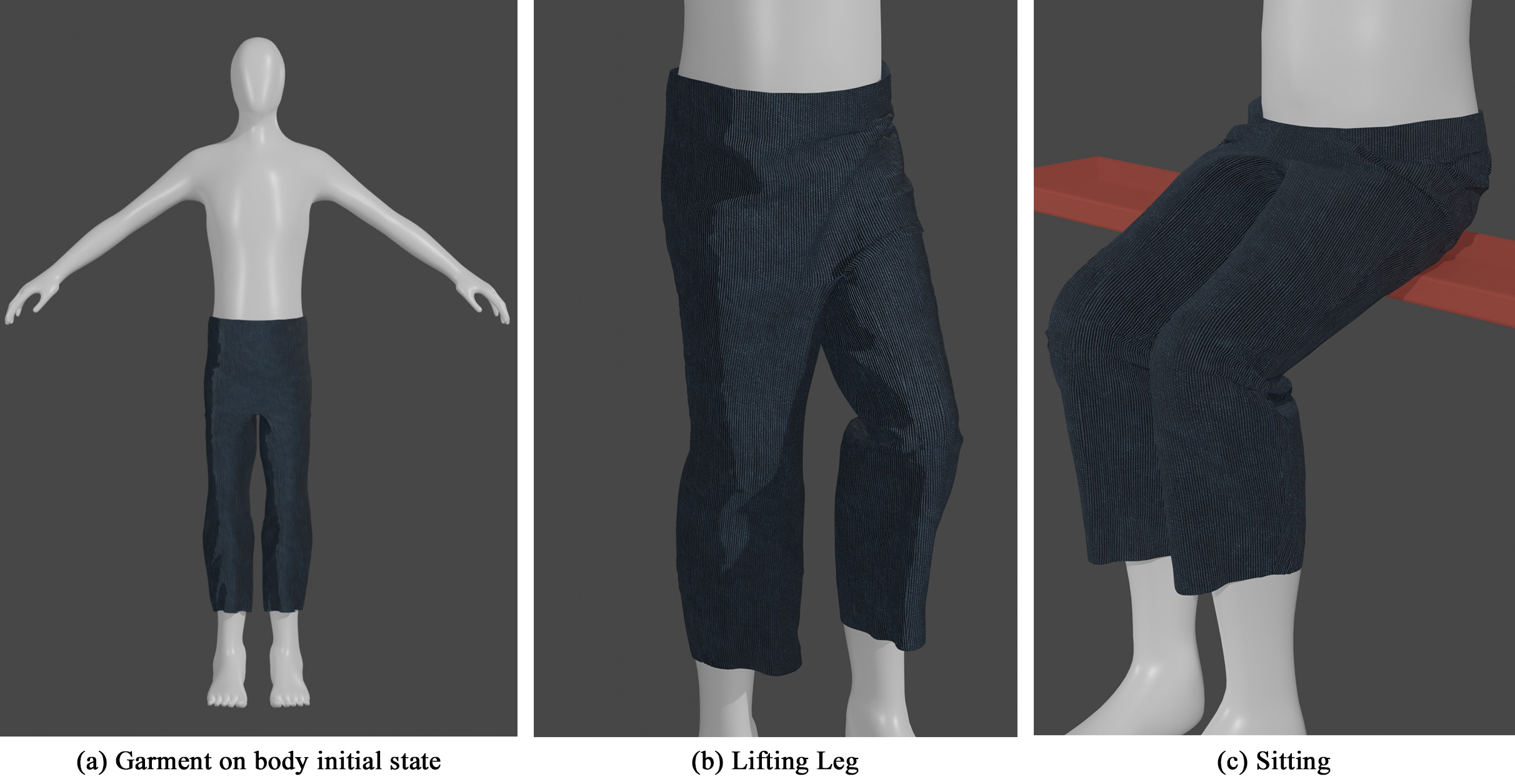}
    \caption{Trousers simulation. (a) The trousers on A-pose human body does not have wrinkles; (b) Lifting a leg deforms the trousers moderately; (c) Sitting down causes larger deformations.}
    \label{fig:motion_garment}
\end{figure}
\begin{figure}[h]
    \centering
    \includegraphics[width=0.48\textwidth]{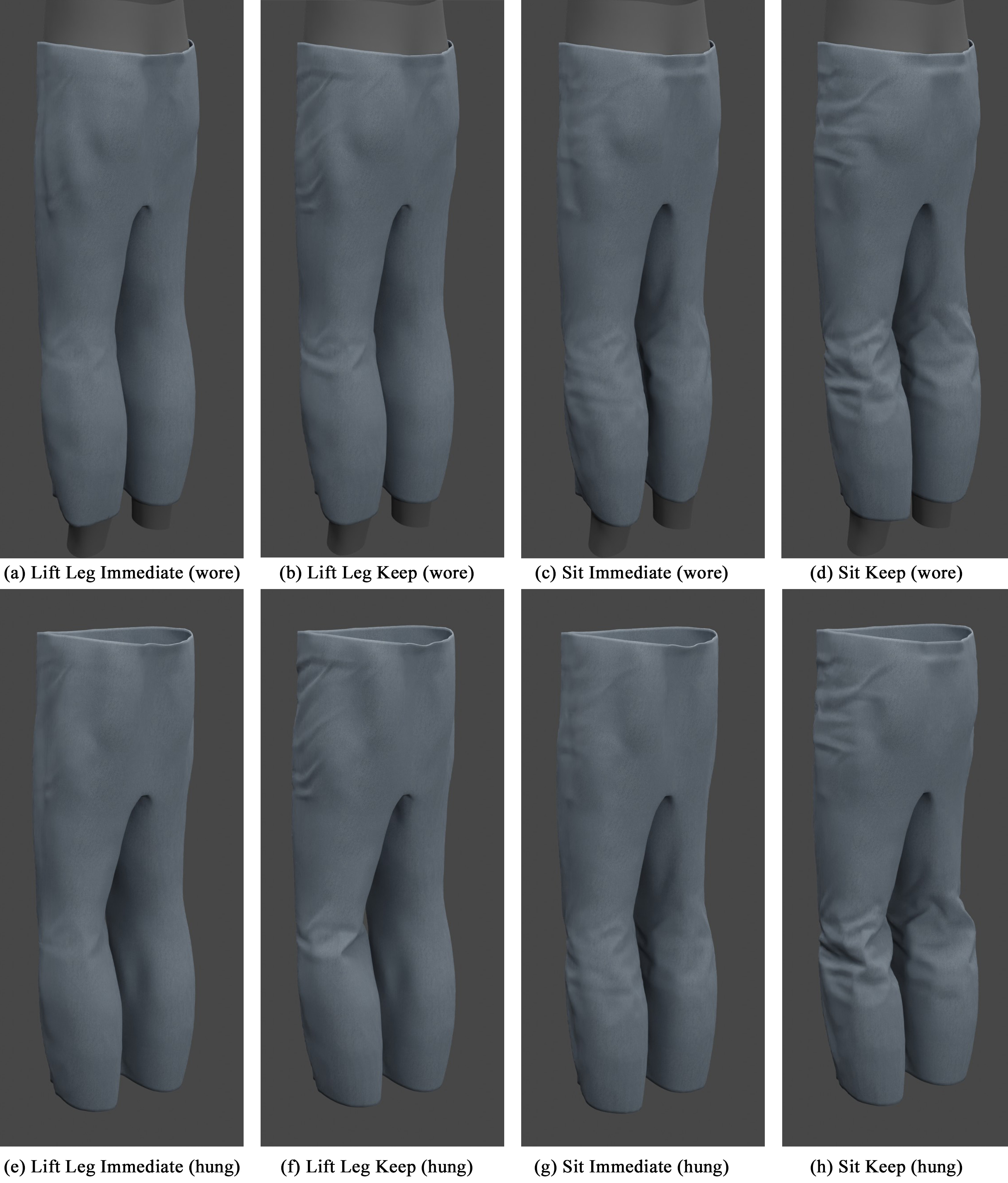}
    \caption{ (a-d) and (e-h) are with and without human body. The wrinkles caused by sitting (c, d, g, h) are more obvious that those caused by lifting leg (a, b, e, f) because sitting causes larger deformations. Moreover, the wrinkles on (b, d, f, h) are sharper and deeper than those on (a, c, e, g). Therefore, keeping deformations for 500s makes the wrinkles more obvious.}
    \label{fig:wrinkle_garment}
\end{figure}

\begin{figure*}
    \centering
    \includegraphics[width=\textwidth]{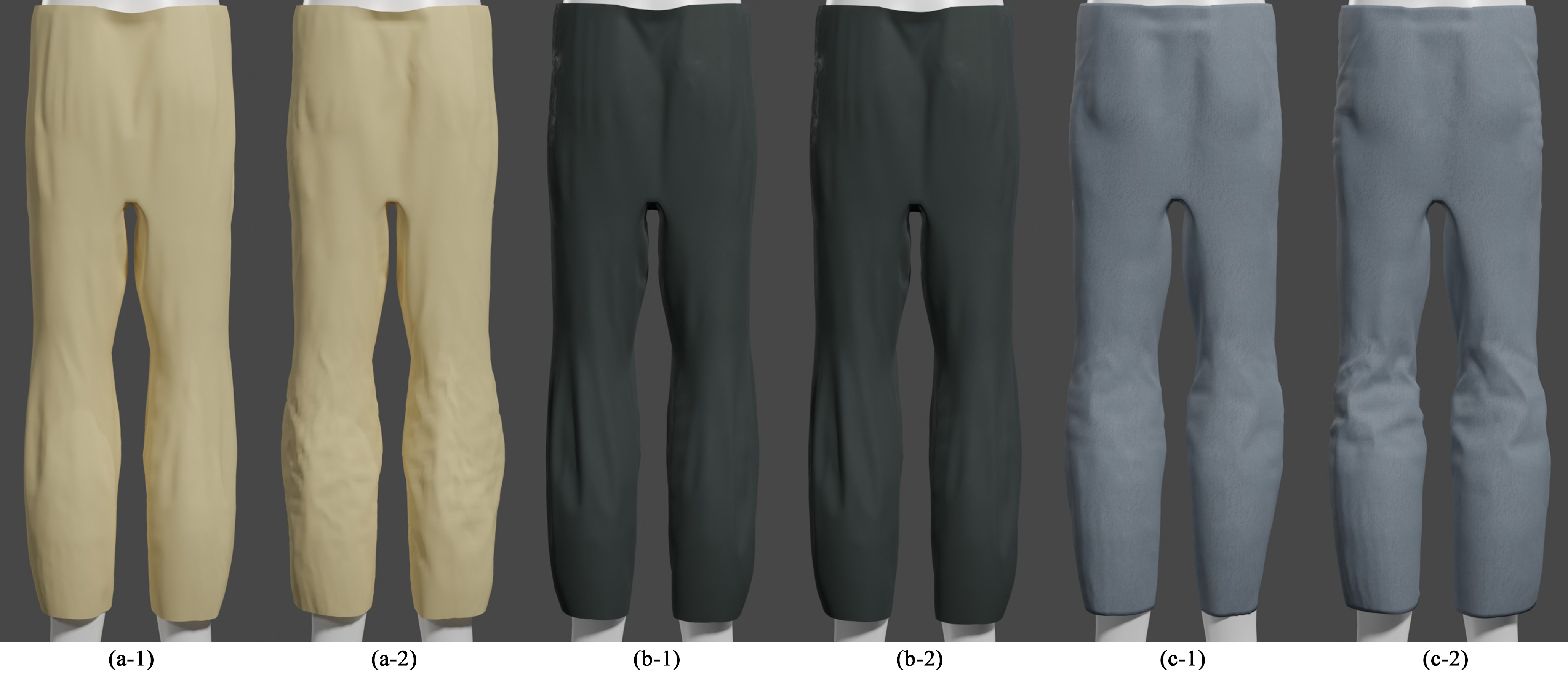}
    \caption{Wrinkles with time-dependency. After sitting for different duration (short: x-1; long: x-2), the wrinkles appear on the trousers made from three types of fabrics: Cotton (a-x), Polyester (b-x), and Denim (c-x). Cotton is soft and thin. Sitting causes small wrinkles (a-1) which are more obvious after sitting for a long time (a-2). Polyester has little plasticity and rarely generates hard wrinkles. Therefore, the wrinkles on the grey polyester trousers are less noticeable (b-1 \& b-2). Denim is stiff and thick. Sitting causes big folds (c-1) which becomes more obvious after sitting for a long time (c-2).}
    \label{fig:trousers_wrinkles}
\end{figure*}

\begin{figure}[h]
    \centering
    \includegraphics[width=0.48\textwidth]{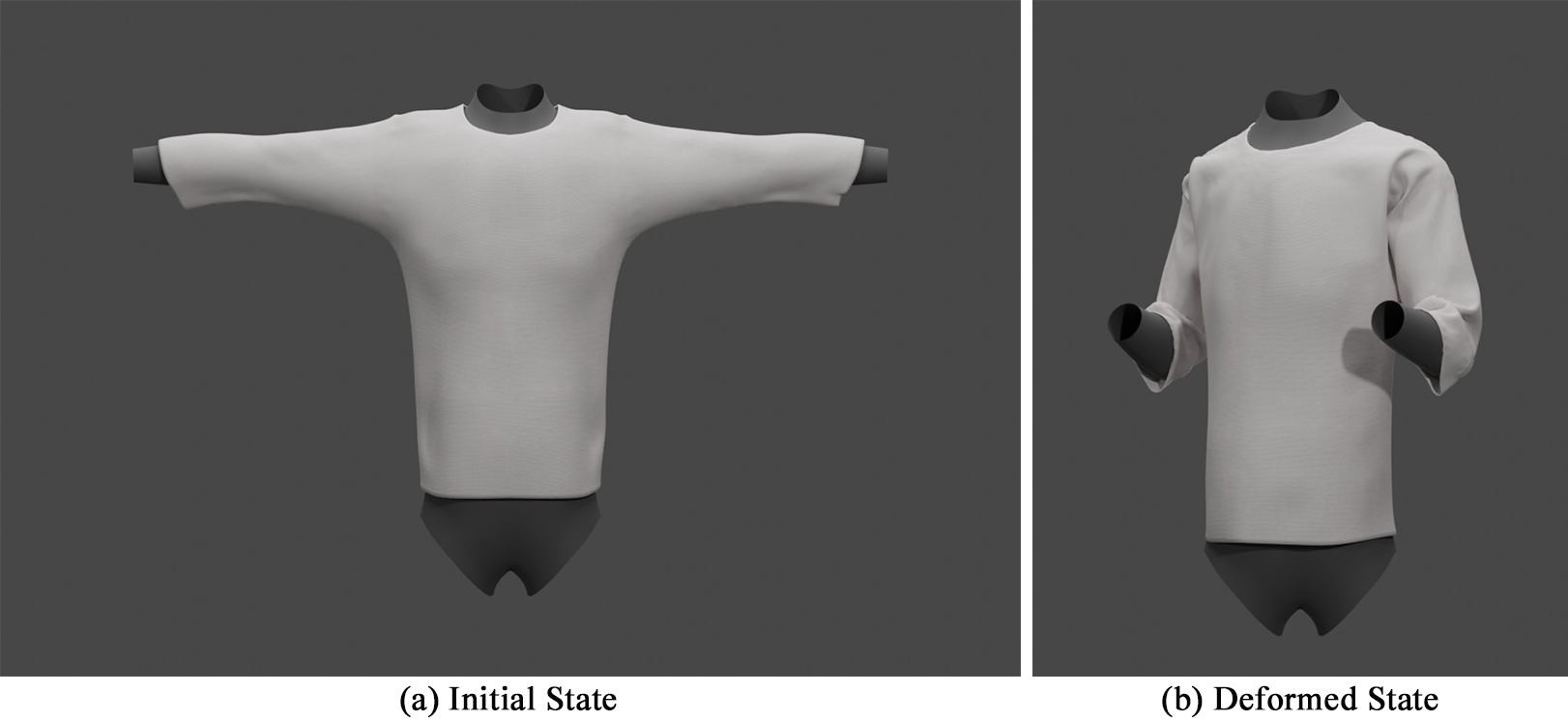}
    \caption{Top garment simulation.(a) The t-shirt on a T-pose human body does not have wrinkles; (b) Bending the arm folds the elbow areas and the underarm area of the sleeves.}
    \label{fig:top_garment_motion}
\end{figure}
\begin{figure}[h]
    \centering
    \includegraphics[width=0.48\textwidth]{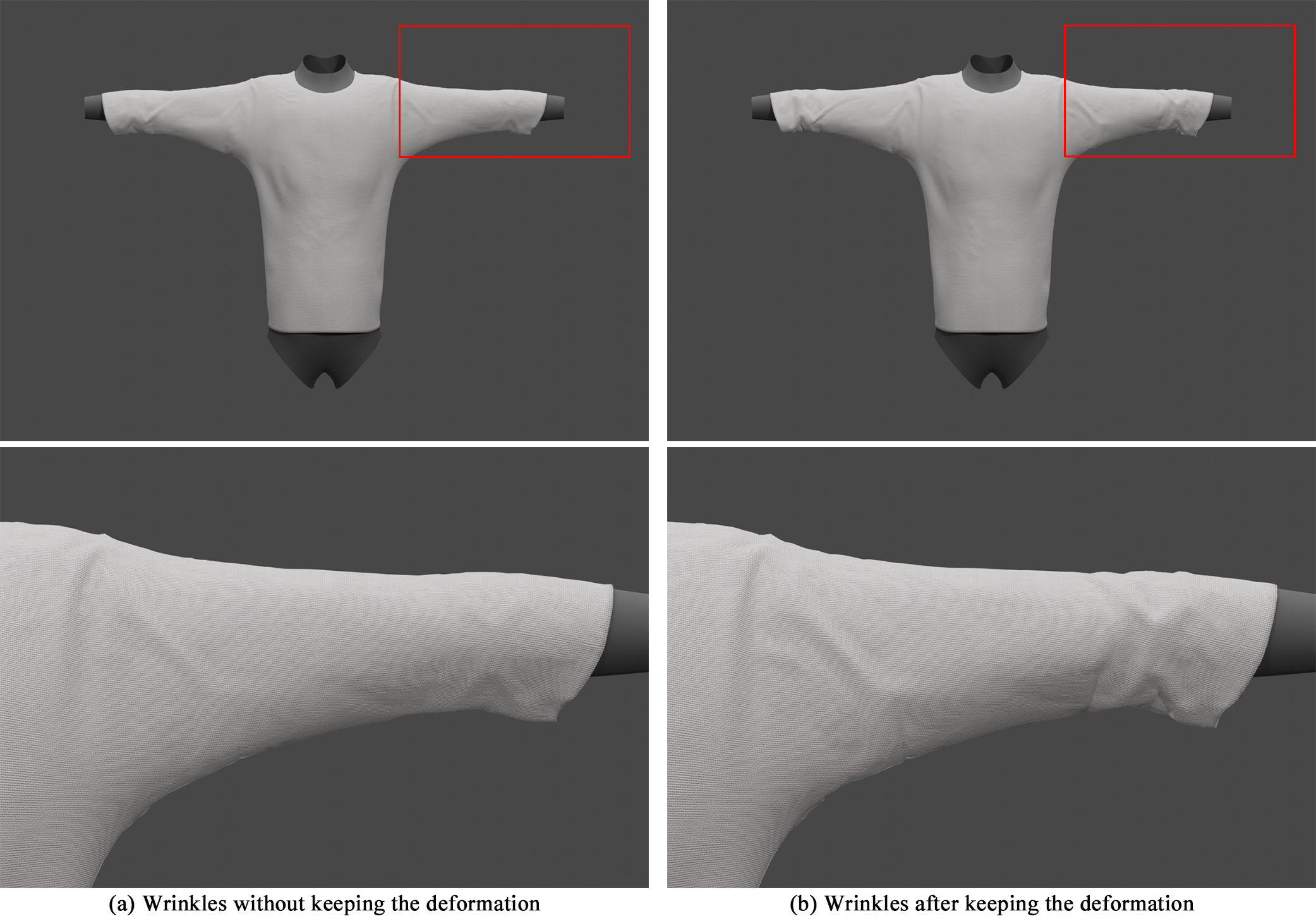}
    \caption{The wrinkles on the t-shirt after returning to the initial t-pose. Keeping the pose for 500s makes the wrinkles (b) more obvious that those formed by immediately returning to the t-pose (a).}
    \label{fig:top_garment_compare}
\end{figure}

\textbf{Compress Wrinkles} We further test our simulator with complex wrinkle patterns caused by multiple forces. We first drop a wrinkleless cloth into a cylindrical container (\Cref{fig:motion_fall} (a)) and let the cloth to fold by gravity and collision (\Cref{fig:motion_fall} (b)). Then our experiment bifurcates into two scenarios. In the first scenario, we remove the container and lift the cloth by picking up its two corners. In the second scenario, we remove the container then add a heavy weight (the transparent object in \Cref{fig:motion_fall} (c)) and let it fall onto the cloth. Within each scenario, to show the time dependence, we either immediately lift the cloth or keep it for 500s. After lifting, the cloth will be hung only under the influence of its weight. The experiment aims to mimic everyday scenarios e.g., clothes dropped onto a sofa and left there, sometimes with people sitting on them. The experiment is designed to see if our method can simulate wrinkles that qualitatively agree with: (1) the measurement in~\cite{levison1962some} stating that cloth persistent wrinkles are time-dependent; (2) the discoveries in~\cite{prevorsek1975influence} stating that persistent wrinkles are collectively caused by internal friction and plastic deformations.

\begin{figure}[tb]
    \centering
    \includegraphics[width=0.48\textwidth]{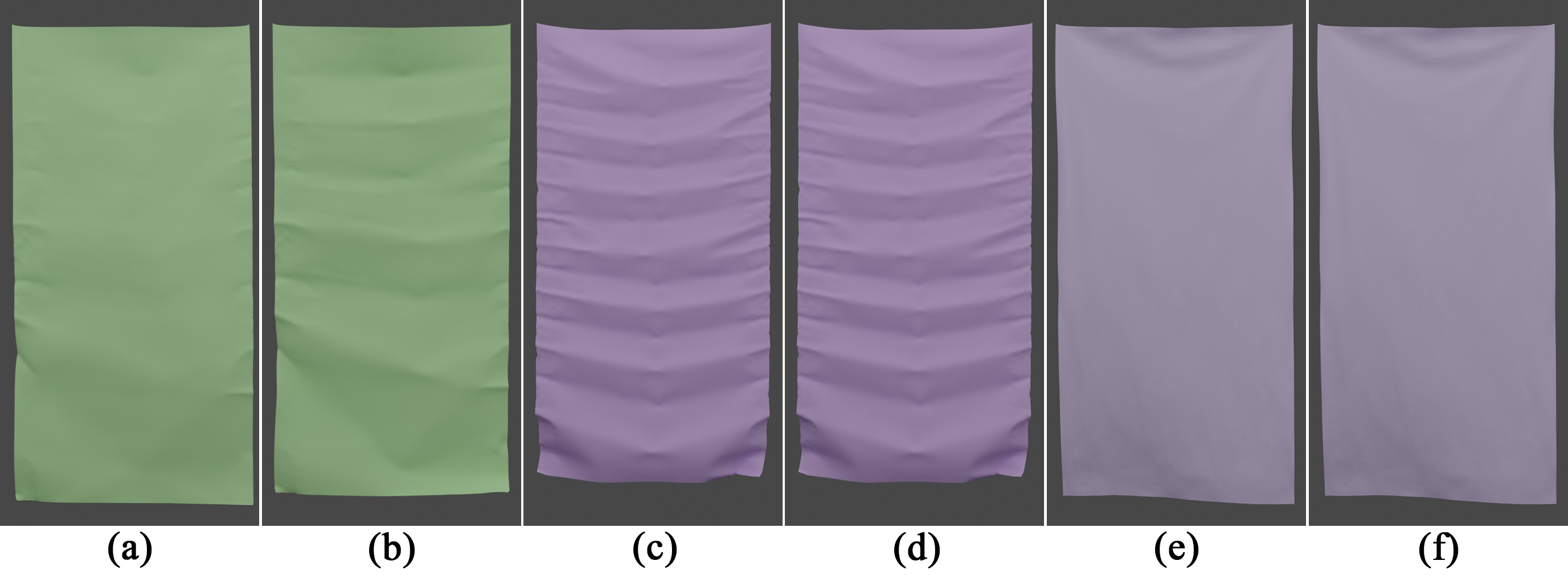}
    \caption{The friction wrinkles simulated by the Dahl's friction model \cite{miguel2013modeling} with different deformation durations: (c) lift immediately; (d) lift after long-time keeping. Due to lacking of time dependence, the wrinkles simulated by the Dahl's model do not vary with time. By contrast, our simulator can simulate time-dependent wrinkles: (a) lift immediately; (b) lift after long-time keeping. (e, f): Owing to lacking of stick friction, the wrinkles simulated by the Dahl's model tend to disappear without slowing cloth motions and using large friction coefficient. A larger figure in SM.}
    \label{fig:dahl_wrinkles}
\end{figure}

In \Cref{fig:cloth_wrinkles}, we first notice that all wrinkles are not reversible by gravity. Looking closely, \Cref{fig:cloth_wrinkles} (c, d) generate sharper wrinkles than \Cref{fig:cloth_wrinkles} (a, b). This is understandable and intuitive because there is no heavy weight placed on the cloth in \Cref{fig:cloth_wrinkles} (a, b). The weight forces larger deformation and therefore makes plasticity prominent in wrinkles. Further, soft wrinkles are visible in all cases, demonstrating the effect of friction. Overall, combining the internal friction and the plasticity gives a visually realistic combination of both soft and hard wrinkles.

\textbf{Time dependence} Next, within each scenario (with and without weight), we can clearly see the time dependence. When there is no weight and the cloth is immediately lifted (\Cref{fig:cloth_wrinkles} (a)), no many wrinkles are visible, as the lifting and hanging leads to stretching under self weight, reversing some soft wrinkles caused by the internal friction. Comparatively, after the cloth is held for 500s (\Cref{fig:cloth_wrinkles} (b)), the friction wrinkles start to harden. Wrinkles become more visible and resist recovery. Note that most of the hardened wrinkles here are still friction wrinkles. Similarly, when there is a weight, even for a short period of time, plastic wrinkles are still formed (\Cref{fig:cloth_wrinkles} (c)). But since the time is short, some of the wrinkles are partially recovered after lifting before they harden completely. If the weight presses a while, then sharp and irreversible wrinkles appear (\Cref{fig:cloth_wrinkles} (d)). Overall, this experiment shows both the internal friction and the plasticity, when combined, are still time-dependent and can generate highly plausible wrinkles.

\textbf{Friction vs Plasticity} Further, comparing \Cref{fig:cloth_wrinkles} (b) and (c), we see steeply sharp wrinkles with high curvature at the peak of the wrinkle ridges in \Cref{fig:cloth_wrinkles} (c) which are not present in \Cref{fig:cloth_wrinkles} (b). This suggests that the internal friction, even after dwell, is less likely to form wrinkles as sharp as plastic wrinkles~\cite{prevorsek1975influence}. This is physically plausible and visually intuitive. The combination of the internal friction and the plasticity enables the simulator to respond to all kinds of deformations and external forces realistically.

\textbf{Tensile Friction and Plasticity} Our model can also simulate wrinkles mainly caused by tensile deformation. We design an experiment where mainly in-plane elongation is induced. We fix the four edges of a cloth and press the central area downwards to cause small and large deformation (\Cref{fig:tensile} (a) and (b)). Again, we then either release it immediately (\Cref{fig:tensile} (c, d)) or keep the deformation for a while (\Cref{fig:tensile} (e, f)), then compare the wrinkles. With small deformation, the wrinkles are mainly caused by the friction; otherwise the plasticity. Again, the friction wrinkles (\Cref{fig:tensile} (c, e)) are not as sharp as the plastic wrinkles (\Cref{fig:tensile} (d, e)), regardless whether the deformation is kept or not. Furthermore, within each type of wrinkles, the longer the deformation is kept, the harder it is for the wrinkles to recover (\Cref{fig:tensile} (g, h)). The results demonstrate our simulator is also effective in simulating wrinkles induced by tensile deformation.

\subsection{Evaluation on Garments} 


\begin{figure}
    \centering
    \includegraphics[width=0.48\textwidth]{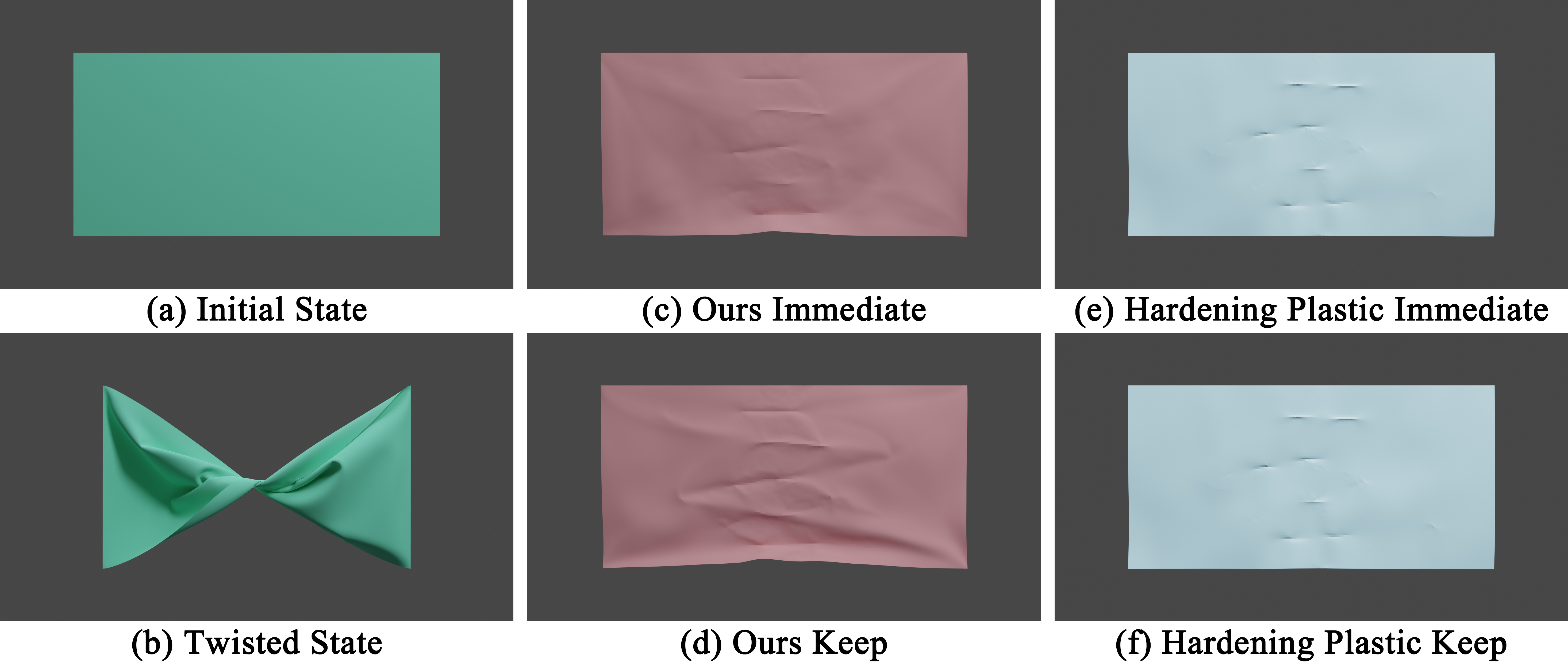}
    \caption{We twist the two edges of a initially flat rectangular cloth (a) in opposite direction by $\frac{\pi}{2}$ to cause wrinkles (b). After flattening the twisted cloth, the wrinkles simulated by our model (pink) and the hardening plastic model (blue) \cite{narain2013folding} with different deformation duration: (c, d) Immediately flattened after twisting; (e, f) Twisted for a while before flattening. A larger figure in SM.}
    \label{fig:comp_plastic}
\end{figure}
\begin{figure*}
    \centering
    \includegraphics[width=\textwidth]{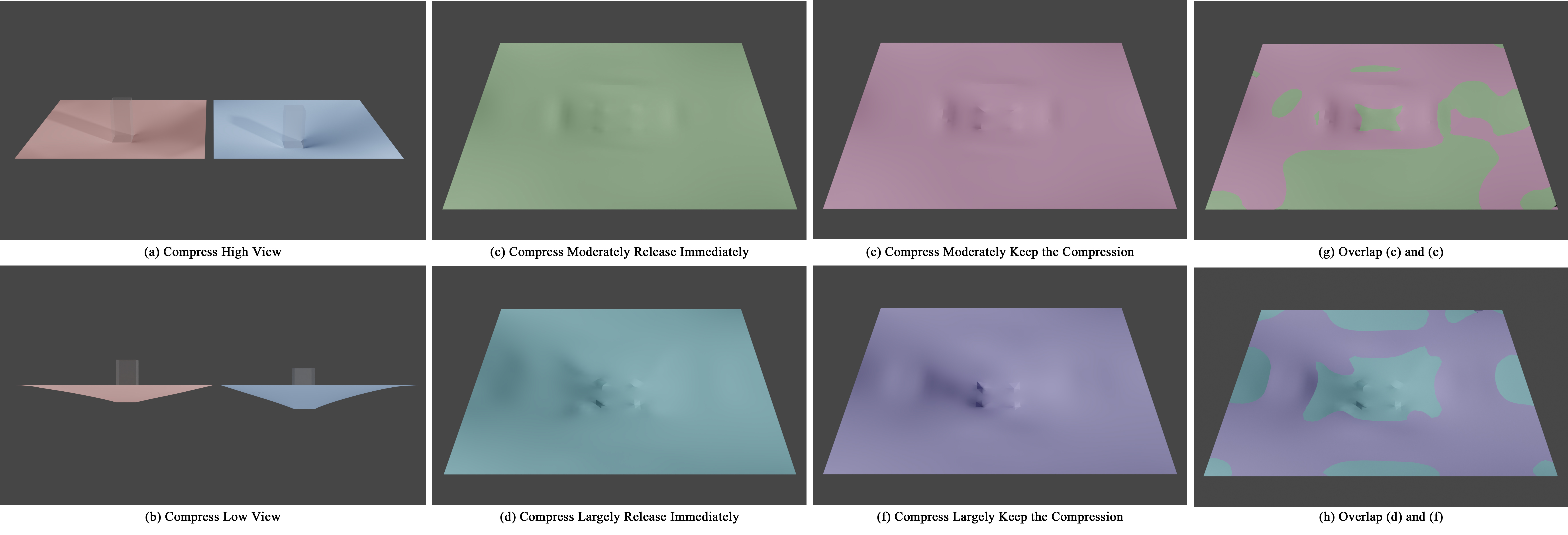}
    \caption{We press a square area of two cloths, whose edges are fixed, to different depth (shown in (a) and (b)). The cloths are persistently stretched due to the stretching friction and plastic deformation (shown in (c-f)). Moreover, (g) and (h) show that pressing the cloth for a long time makes the persistent stretching deformation more obvious.}
    \label{fig:tensile}
\end{figure*}

Our cloth simulation can be easily generalized to garments made from different fabrics with human motions, which is crucial applications from fashion design to animation. We simulate trousers with 20k vertices on a human body (\Cref{fig:motion_garment} (a)), under deformations caused by two common motions: lifting the left leg (\Cref{fig:motion_garment} (b)) and sitting down (\Cref{fig:motion_garment} (c)). Lifting leg is a good example of moderate deformations asymmetrically distributed between two legs, while sitting down involves multiple regions of large deformations around the pelvis and knees.

We first show results in denim trousers in \Cref{fig:wrinkle_garment,fig:teaser}. When simulating the trousers, we keep the lifting or sitting pose for a short duration and 500s. The leg lifting mainly causes wrinkles behind the left knee, with some mild wrinkles behind the bottom. As expected, the wrinkles after the long deformation (\Cref{fig:wrinkle_garment} (b)) are sharper than (\Cref{fig:wrinkle_garment} (a)). Since leg lifting causes moderate wrinkles, the visual difference between short and long deformations is visible but subtle. Comparatively, the difference is more visible in sitting where sharper wrinkles are formed in \Cref{fig:wrinkle_garment} (d) than \Cref{fig:wrinkle_garment} (c). The differences are especially noticeable not only in the areas behind the bottom and around the knees, but also near the crotch. Compared with the real world observations shown in~\Cref{fig:teaser} and Figure 1 in SM, our simulations are visually plausible in the time dependence and the varied wrinkles caused by different motions.

Next, we show the trousers without the human body in \Cref{fig:wrinkle_garment} (e-h). This is to show some wrinkles are generated and maintained mainly due to the collisions between the body and the trousers. Without the human body, the trousers are more stretched under gravity and some wrinkles disappear. However, these disappeared wrinkles tend to be the ones that are soft and mainly caused by the internal friction. These wrinkles are more easily reversible. We also show trousers made from different materials from behind when draping in \Cref{fig:trousers_wrinkles}. Further, we show another example of the wrinkles on a T-shirt in~\Cref{fig:top_garment_motion}, where the simulated wrinkles also exhibit visually convincing time dependence (shown in~\Cref{fig:top_garment_compare}).


\subsection{Comparison Experiments}

To the best of our knowledge, there is no cloth simulator that combines internal friction and plasticity. Further, there is no cloth simulator that models wrinkle time dependence. So we choose two baseline methods that are closest to ours. One only considers the internal friction with using Dahl's Model~\cite{miguel2013modeling}. The other considers plasticity only~\cite{narain2013folding, gingold2004discrete}. Note neither models time dependence.

\textbf{Compare with Dahl's Model} \cite{miguel2013modeling} introduces a Dahl's friction model to simulate cloth internal friction. They show that their friction model can also form cloth wrinkles. However, without considering the dwell effect, it cannot simulate time-dependent friction wrinkles. We use the simulation scenario in \Cref{fig:motion_fall} (b) (without weight) and use Dahl's friction model to simulate the friction wrinkles. \Cref{fig:dahl_wrinkles} (c) and (d) show, when using Dahl's friction model, the wrinkles are identical even though their deformation duration is different. Therefore, Dahl's friction model cannot simulate time-dependent friction wrinkles. In addition, due to the lack of stick friction, Dahl's friction model needs damping to prevent persistent wrinkles from disappearing because of vibrations or fast motions~\cite{miguel2013modeling}, as shown in \Cref{fig:dahl_wrinkles} (e, f). To simulate the obvious wrinkles as shown in \Cref{fig:dahl_wrinkles} (c, d), we use large friction coefficient and frequently slow down motion by setting velocity to zero to keep wrinkles stable. Thus, compared with our model, it is more difficult to use to simulate persistent wrinkles.

\textbf{Compare with Hardening Plastic Model.} Plastic models are commonly used for simulating cloth wrinkles in graphics. We choose the hardening plastic model as the baseline which has been used in \cite{narain2013folding, gingold2004discrete}. We twist the two edges of a rectangular cloth to cause plastic deformations (\Cref{fig:comp_plastic} (a-b)). Our model and the baseline use the same yield strain, $\varepsilon_Y$. As shown in \Cref{fig:comp_plastic} (c) and (d), keeping the deformation for different durations affects the wrinkles when using our model. Conversely, as shown in \Cref{fig:comp_plastic} (e) and (f), the wrinkles simulated by the baseline method are identical no matter how long the deformation is kept. Moreover, due to the lack of internal friction, the cloths in \Cref{fig:comp_plastic} (e) and (f) have much fewer wrinkles because plastic wrinkles only appears in large deformations.

\section{Limitations}

The biggest limitation is that it is hard to quantitatively evaluate the simulation results against real-world cloths. This is because specialized equipment with precision is needed for compression/twisting tests, where wrinkles form in the presence of buckling. However, even in the absence of quantitative evaluations, we argue, as an animation tool, our simulator still produces highly believable results. In addition, similar to other simulators, our method currently relies on hand-tuning for simulating different materials. So it is difficult to automatically calibrate the model using data. As other simulators, a more fine-grained model does need more parameters. For instance, Maya’s uses 9 parameters to simulate isotropic cloths. Marvelous Designer uses 15 parameters to introduce anisotropy. Blender uses 20+ parameters for more effects. We believe it is acceptable that adding 8 parameters for simulating an important visual effect common in real cloth wrinkles. To ease parameter tuning, we provide a parameter tuning guide in the SM to further explain the functionalities of each parameters. Finally, by using the anchor-based method, although our friction model is simple and straightforward, it may have stability issues when adopting a large time step size, we discuss and analysis it in the SM.
\begin{figure*}
    \centering
    \includegraphics[width=\textwidth]{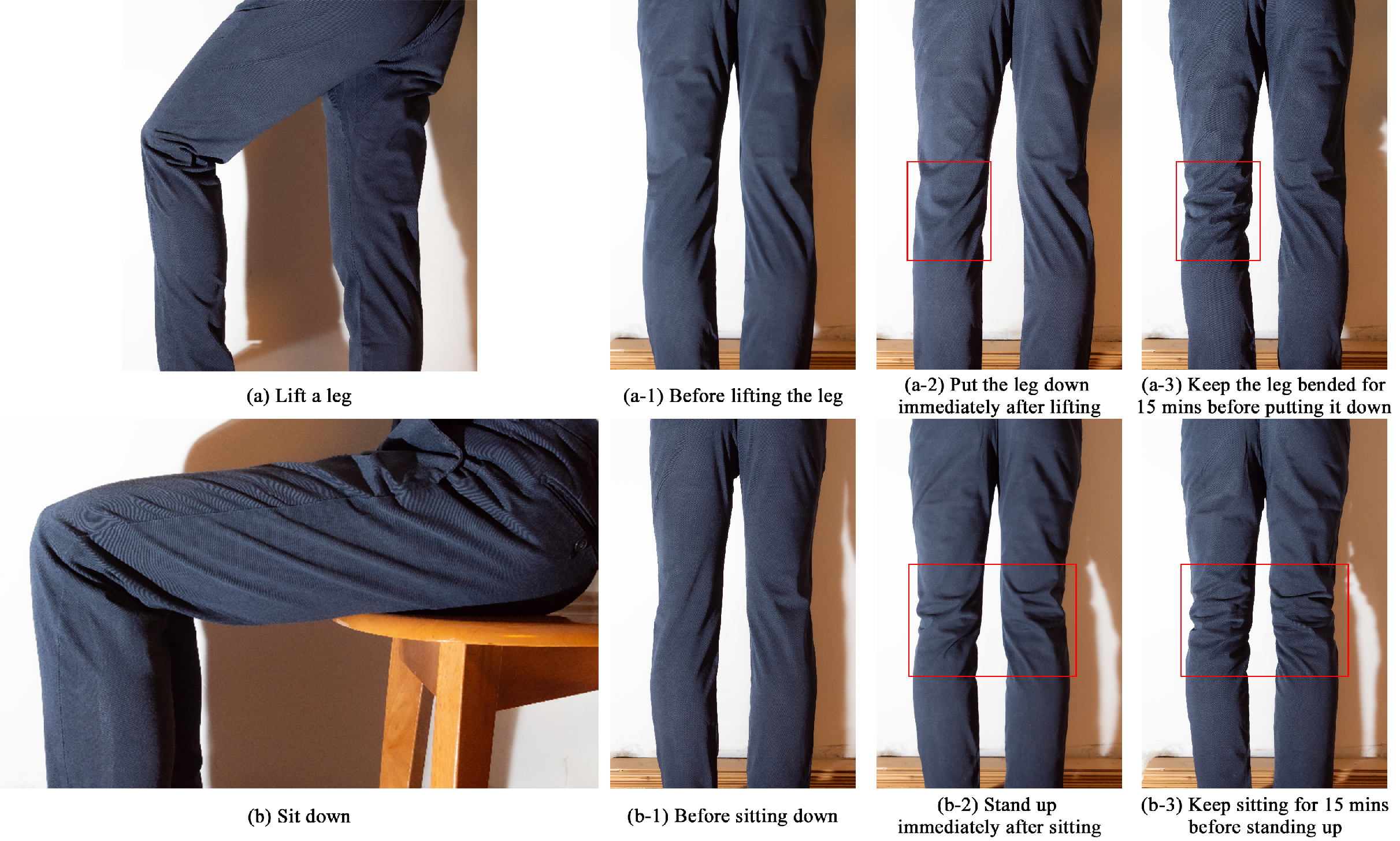}
    \caption{The wrinkles observed on a pair of real trousers. Lifting a leg (a) and sitting down (b) bend the knee areas of the trousers and thus causes wrinkles (a-2) and (b-2). The wrinkles results from sitting down are more sharper than those caused by lifting leg (a-2) because the former pose deforms the trousers more extremely. Moreover, keeping either pose for a while (15 mins) tends to make formed wrinkles more solid, shown in (a-3) and (b-3).}
    \label{fig:real_trousers}
\end{figure*}
\section{Discussion and Conclusion}

We have introduced a new cloth simulator that is capable of generating visually realistic persistent wrinkles. To the best of our knowledge, our method is the first one that investigates the temporal aspect of wrinkle formation. Although it is widely studied in textiles and regarded as crucial in capturing clothes’ mechanics and appearance~\cite{levison1962some}, relevant studies are still absent in graphics. To this end, we proposed a new physics-inspired model combining time-dependent internal friction and plasticity. Through experiments, we have demonstrated high visual plausibility by introducing complex cloth wrinkle mechanics: (1) the interplay between internal friction and plastic deformation; (2) their dependence on time.

To our knowledge, none of the existing cloth simulator can simulate the time dependence easily. The existing physics-based, data-driven, and self-supervised cloth simulators cannot simulate time dependence. Admittedly, rule-based methods can achieve this, but would be prohibitively time-consuming because, to simulate time dependence, they need to define the wrinkles everywhere on the garment for any time duration of deformation of any magnitude. Conversely, our method does not require these laborious manual works once the physical parameters are determined. In addition, the parameters in our model embed physical meaning which are easier to understand and tweak than those in rule-based methods. Our work introduces a new physical temporal effect to graphics simulation, and can be very useful in animations,  visual effects. This work can also meet the needs from the fashion designers who expect to view garments' wrinkles at the design stage.

Additionally, our work also the first time models the interplay between internal friction and plasticity. Although they can produce similar behaviors but are different in essence, and either is indispensable. In fabrics, inter-yarn friction is not enough to fully model a cloth's plasticity. For example, plastic creeping is commonly observed in real clothes~\cite{jung2016modeling}, which is impossible to simulate only by the friction model, but our time-dependent plastic model can simulate it. On the other hand, only using plastic model is not enough either because it cannot simulate the hysteresis in moderate deformations observed in the real load-deformation curves~\cite{miguel2013modeling}. Moreover, it has been observed that the internal friction in filament yarns is considerably less than it in staple yarns~\cite{huffington1961internal}. Thus, only using internal friction to model yarns' plasticity is insufficient.

In future, we will design more controllable and standardized experiments, such as using LLY-02 or YG541D testing apparatuses, so that we can conduct accurate quantitative evaluations. Further, we will release more parameters for realistic simulations. We will also explore automatically learning these parameters from real cloths and predict real cloths' behaviors. Considering the required large training data and long data collecting time, we leave it to future work. Theoretically, our model can be used in a plug-and-play manner. For example, our friction and plastic models are not bounded to the (hyper)elastic models used in this work and we will explore to combine them with other constitutive models. We will also try to integrate it into other cloth simulators, e.g., yarn-level cloth simulators, by using our friction to model inter-yarn frictions and using our plastic model to model yarns' plastic deformations. Last but not least, we adopt exponential function to model the time dependence based on the real measurements in~\cite{levison1962some}, we will try other alternatives to simulate different fabric materials or do stylized simulations.

\newpage

\appendix

\begin{appendices}

\section{Additional Experiments}

\subsection{More Visual Results}

\begin{figure*}[htb]
    \centering
    \includegraphics[width=\textwidth]{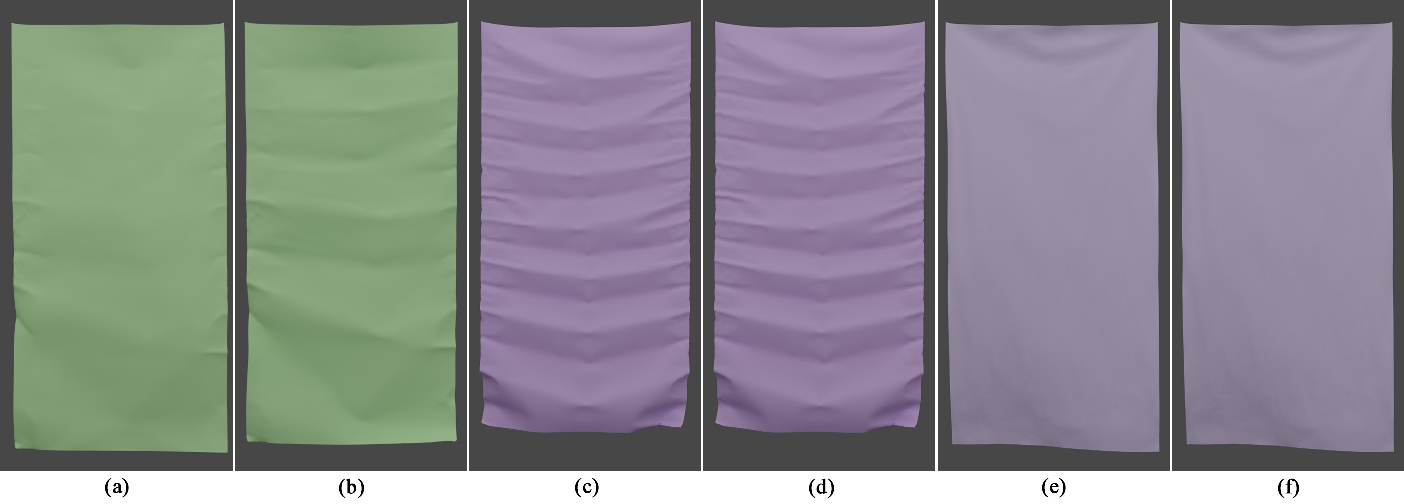}
    \caption{The friction wrinkles simulated by the Dahl's friction model \cite{miguel2013modeling} with different deformation durations: (c) lift immediately; (d) lift after long-time keeping. Due to lacking of time dependence, the wrinkles simulated by the Dahl's model do not vary with time. By contrast, our simulator can simulate time-dependent wrinkles: (a) lift immediately; (b) lift after long-time keeping. (e, f): Owing to lacking of stick friction, the wrinkles simulated by the Dahl's model tend to disappear without slowing cloth motions and using large friction coefficient.}
    \label{fig:dahl_wrinkles_app}
\end{figure*}

\begin{figure*}
    \centering
    \includegraphics[width=\textwidth]{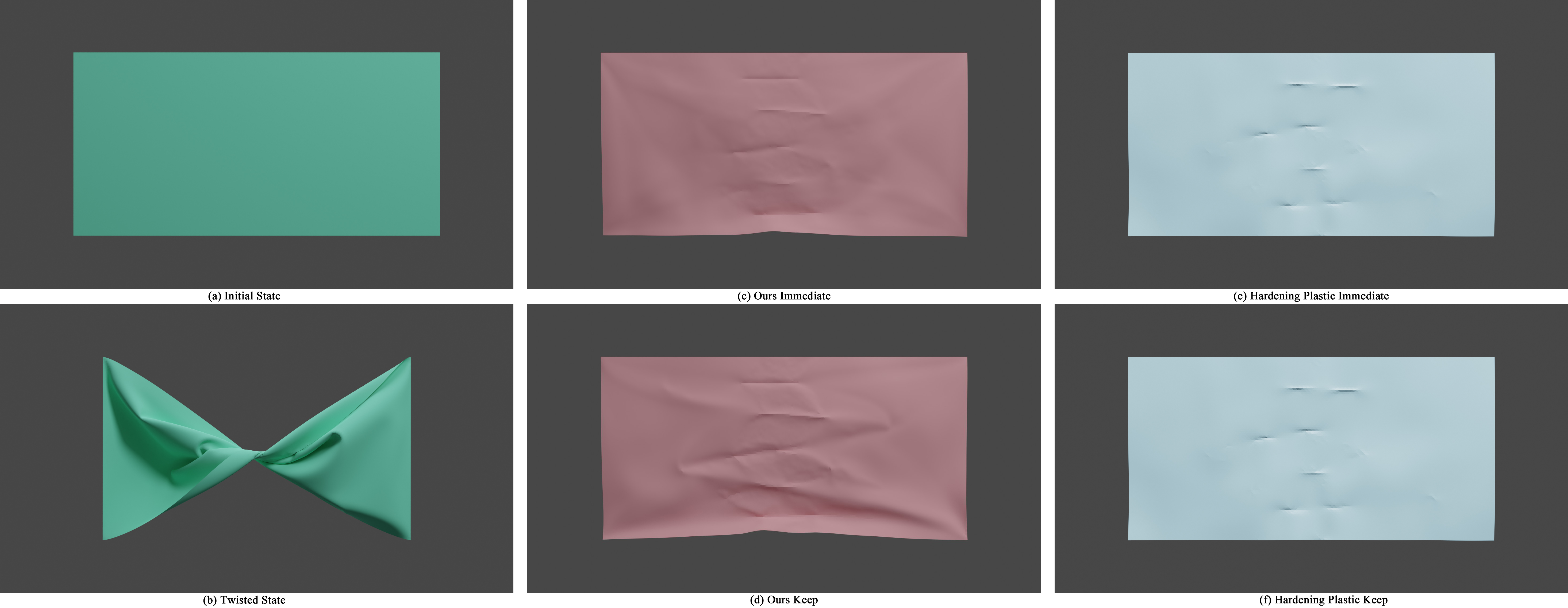}
    \caption{We twist the two edges of a initially flat rectangular cloth (a) in opposite direction by $\frac{\pi}{2}$ to cause wrinkles (b). After flattening the twisted cloth, the wrinkles simulated by our model (pink) and the hardening plastic model (blue) \cite{narain2013folding} with different deformation duration: (c, d) Immediately flattened after twisting; (e, f) Twisted for a while before flattening.}
    \label{fig:comp_plastic_app}
\end{figure*}

\begin{figure}
    \centering
    \includegraphics[width=\linewidth]{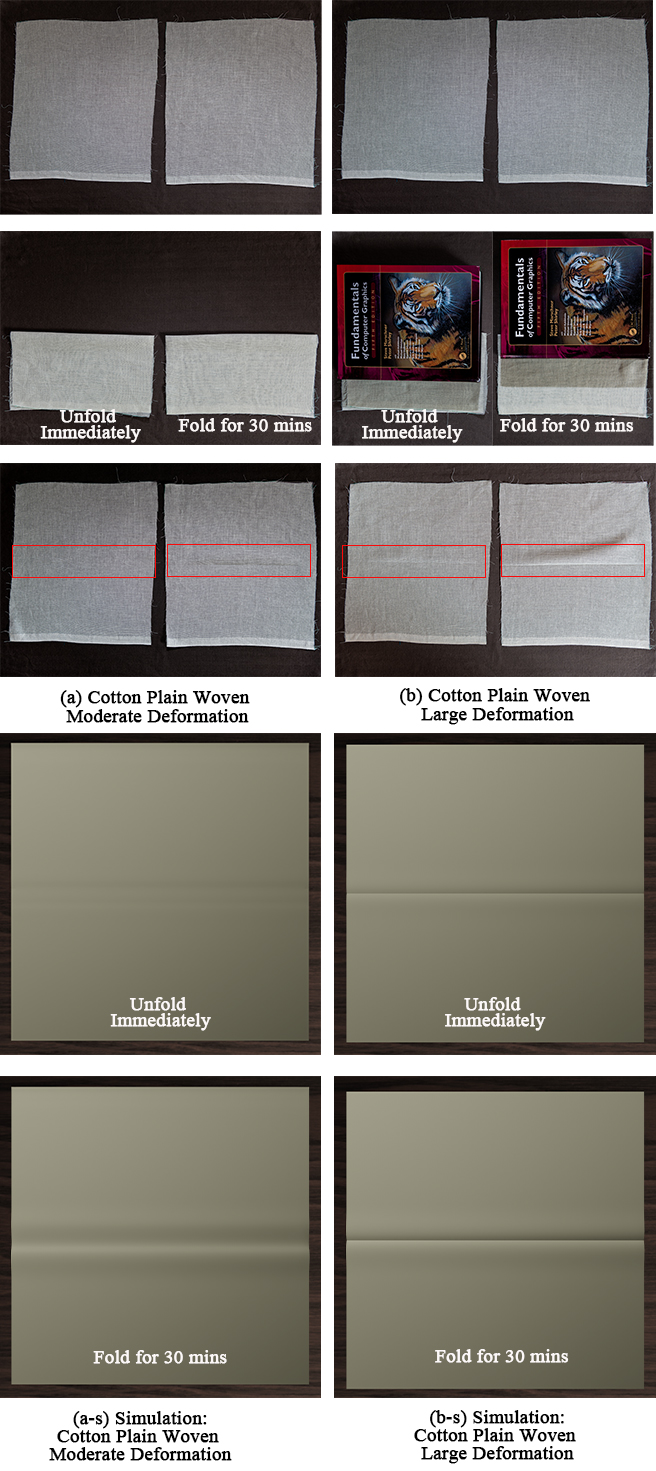}
    \caption{Our simulator can realistically simulate time-dependent wrinkles (a-s, b-s) observed in the real cotton cloth samples (a, b) in different deformations.}
    \label{fig:material_cotton}
\end{figure}

\begin{figure}
    \centering
    \includegraphics[width=\linewidth]{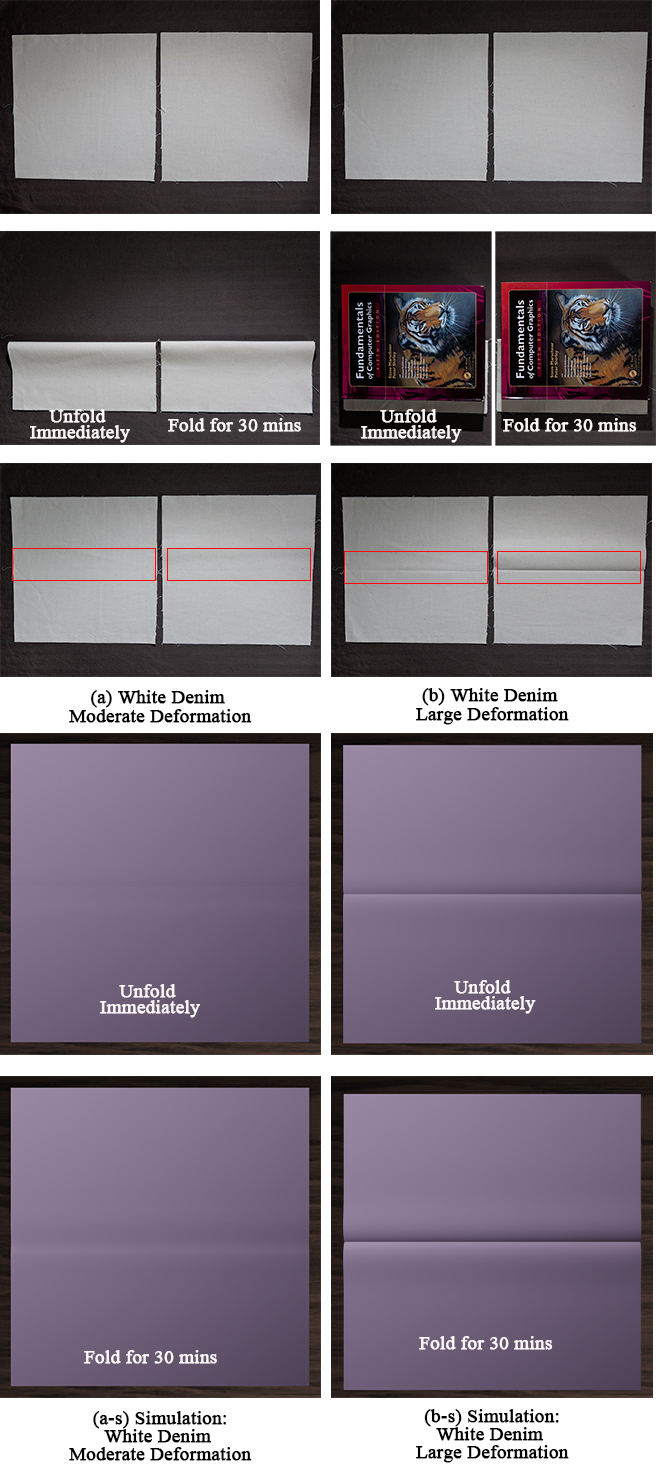}
    \caption{Our simulator can realistically simulate time-dependent wrinkles (a-s, b-s) observed in the real white denim cloth samples (a, b) in different deformations.}
    \label{fig:material_denim}
\end{figure}

\begin{figure}
    \centering
    \includegraphics[width=\linewidth]{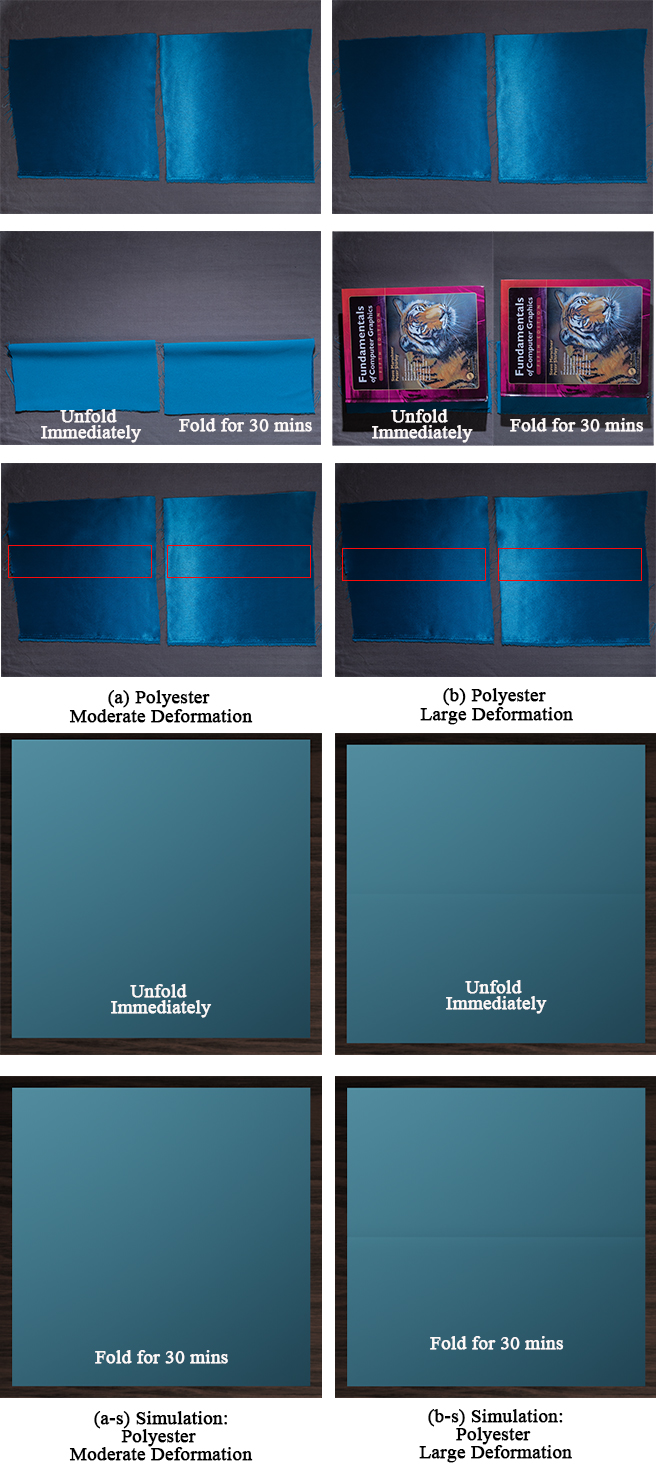}
    \caption{Our simulator can realistically simulate time-dependent wrinkles (a-s, b-s) observed in the real polyester cloth samples (a, b) in different deformations.}
    \label{fig:material_poly}
\end{figure}

\begin{figure*}
    \centering
    \includegraphics[width=\textwidth]{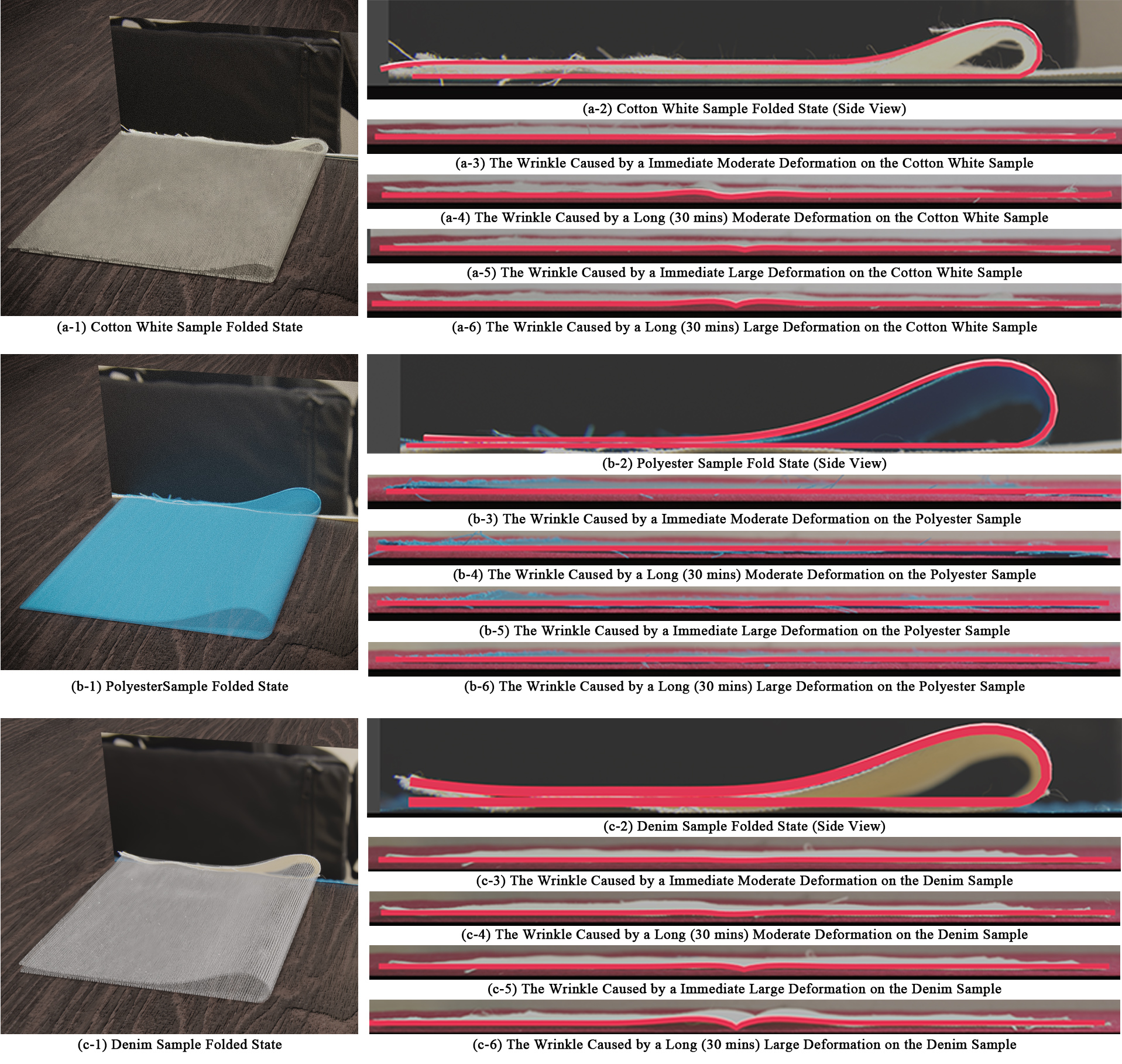}
    \caption{The simulated clothes' folded state closely overlap real counterpart shown in the photos (a-1, b-1, and c-1) where we use red to highlight the clothes from the side view (a-2, b-2, and c-2). In addition, the simulated wrinkles formed in different deformations also closely fit the real clothes (a-[3-6], b-[3-6], and c-[3-6]). The wrinkles formed in Cotton White clothes are mild (a-[3-6]), formed in Polyester are very small (b-[3-6]), and formed in Denim are more obvious (c-[3-6]).In addition, folding or compressing clothes makes the wrinkles are obvious, especially in Cotton White and Denim clothes (a-4, a-6, c-4, c-6).}
    \label{fig:eval_sample_side}
\end{figure*}

\Cref{fig:dahl_wrinkles_app} and \Cref{fig:comp_plastic_app} provide a larger view of \Cref{fig:dahl_wrinkles} and \Cref{fig:comp_plastic} in the main paper respectively. \Cref{fig:material_cotton} and \Cref{fig:material_denim} zoom in the wrinkles in the \Cref{fig:eval_sample} of the main paper. In addition, the wrinkles formed on real polyester cloth samples caused by different deformations and corresponding simulations are shown in~\Cref{fig:material_poly}. \Cref{fig:material_poly} (a,b) show that only the large deformation can result in small wrinkles, and our simulator can plausibly simulate the polyester's material characteristics (a-s, b-s). Moreover, we also compare the side view of the simulated and real cloth wrinkles in~\Cref{fig:eval_sample_side} which shows our simulated wrinkles can closely fit those on the real clothes. Folding or compressing clothes makes the wrinkles sharper, especially obvious in the Cotton White and Denim clothes~\Cref{fig:eval_sample_side} (a-3, a-4, a-5, a-6, c-3, c-4, c-5, c-6.). Conversely, the wrinkles formed on polyester are far smaller (\Cref{fig:eval_sample_side} (b-3, b-4, b-5, b-6)).

\subsection{Twisting}

\begin{figure}[htb]
    \centering
    \includegraphics[width=0.48\textwidth]{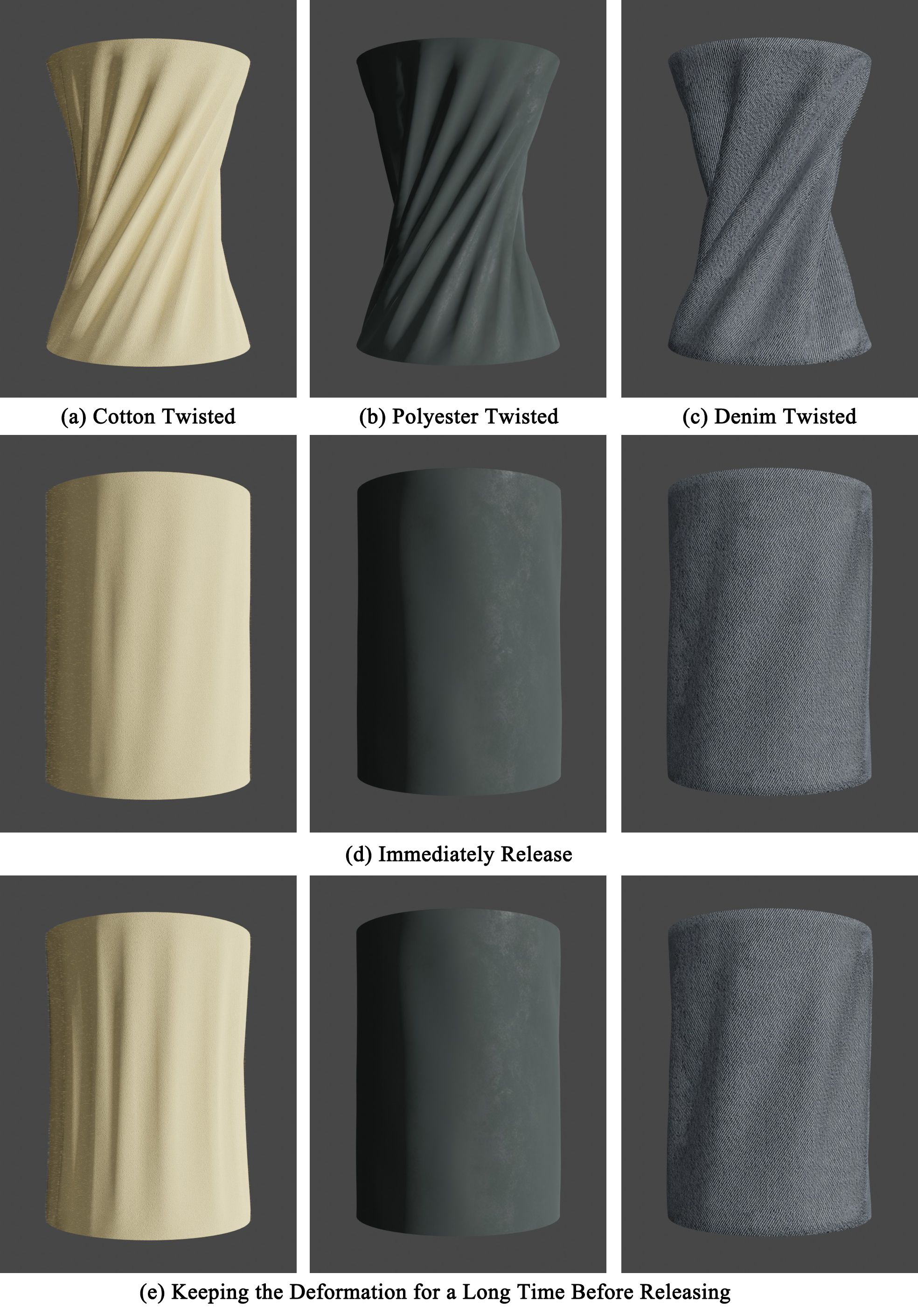}
    \caption{Wrinkles on the twisted cloths made from different materials: (a) Cotton; (b) Denim; (c) Polyester. Cotton and Denim are more likely to generate wrinkles than Polyester. Moreover, keeping the cloths in the twisted state (e) makes the wrinkles more obvious than those on the cloths that are released immediately (d).}
    \label{fig:diff_material_twist}
\end{figure}

To clearly show the persistent wrinkles on the cloths made from different materials, we simulate simple cloth samples of Cotton, Polyester, and Denim in a more controlled scenario. The cloth samples are made into cylinders. We twist the top of every cloth by $0.25\pi$ in the clockwise direction (from the top view) along its central axis with its bottom fixed. Then, we twist in the counter-clockwise direction by $0.25\pi$ to return its initial state and observe the wrinkles caused by the deformation. The results are shown in \Cref{fig:diff_material_twist}. As expected, Cotton gives the most obvious wrinkles and Polyester gives the least. Moreover, keeping the deformation for a long time makes the wrinkles more obvious, which is shown in both Cotton and Denim. Polyester is less affected by deformation duration as there is little internal friction and plasticity. These results conform to the real world observations. Overall, our simulator is capable of simulating high-fidelity wrinkles that reflect cloth materials. 

\subsection{Ablation Study}

\begin{figure*}[htb]
    \centering
    \includegraphics[width=\textwidth]{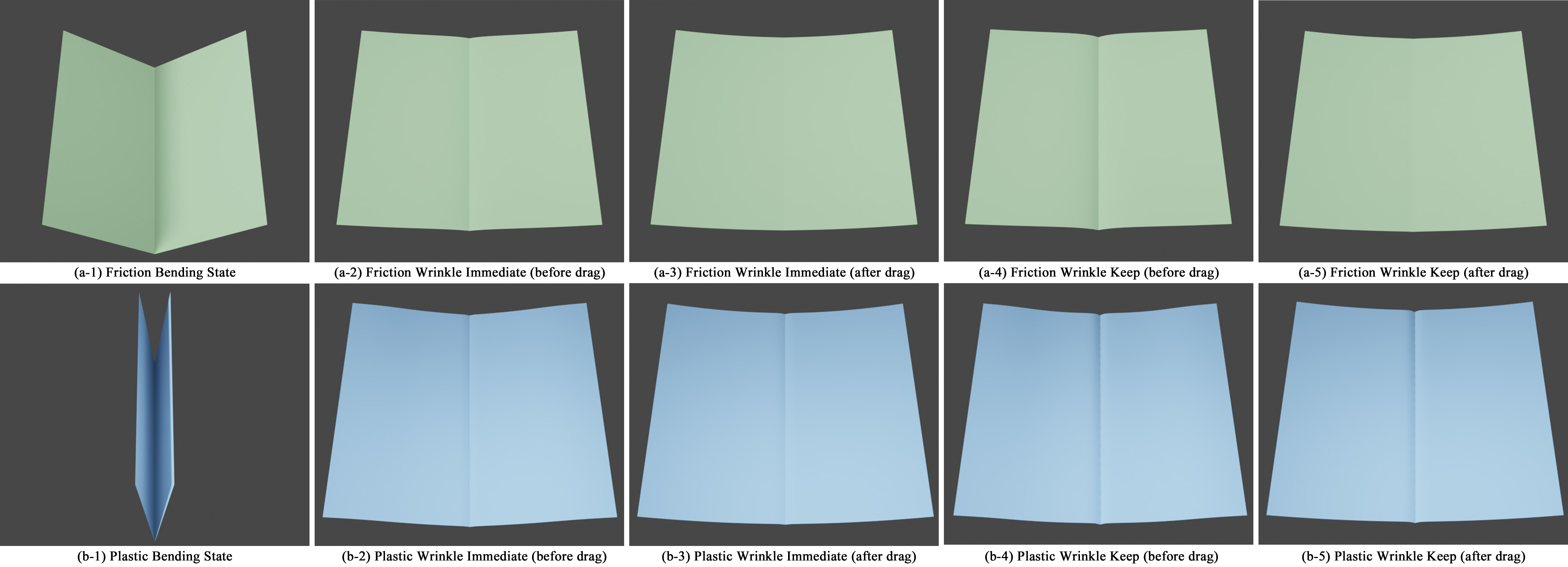}
    \caption{The friction (top) and the plastic (bottom) wrinkles separately simulated by the internal friction model and the plastic model. After the initial bending (a-1, b-1), we either immediately release it (a-2, b-2) or keep it for a while (a-4, b-4), after which we stretch the cloth trying to reverse the wrinkle (a-3, a-5, b-3, b-5). The wrinkles tend to be more recoverable after the immediate release (b, e) than being kept for a while (c, f). Also, plastic wrinkles are harder to recover than friction wrinkles.}
    \label{fig:ablation_friction_plastic}
\end{figure*}

\textbf{Friction/Plastic Wrinkles} To verify that both the internal friction and the plasticity can cause wrinkles, we show them separately by disabling one factor and simulate cloth using the other. We design a simple one-wrinkle experiment shown in \Cref{fig:ablation_friction_plastic}, where a small deformation (\Cref{fig:ablation_friction_plastic} a-1) is used to generate a pure friction wrinkle. After the deformation, we release it either immediately or after keeping the deformation for a long period. After releasing it, we slightly stretch the cloth by pulling the left and the right side in opposite directions, to attempt to flatten the cloth (testing if the wrinkle is reversible). With the immediate release (\Cref{fig:ablation_friction_plastic} a-2 and a-3), the wrinkle is largely reversible; while if the deformation is kept for a while, it is less so (\Cref{fig:ablation_friction_plastic} a-4 and a-5). 

\begin{figure*}[htb]
    \centering
    \includegraphics[width=\linewidth]{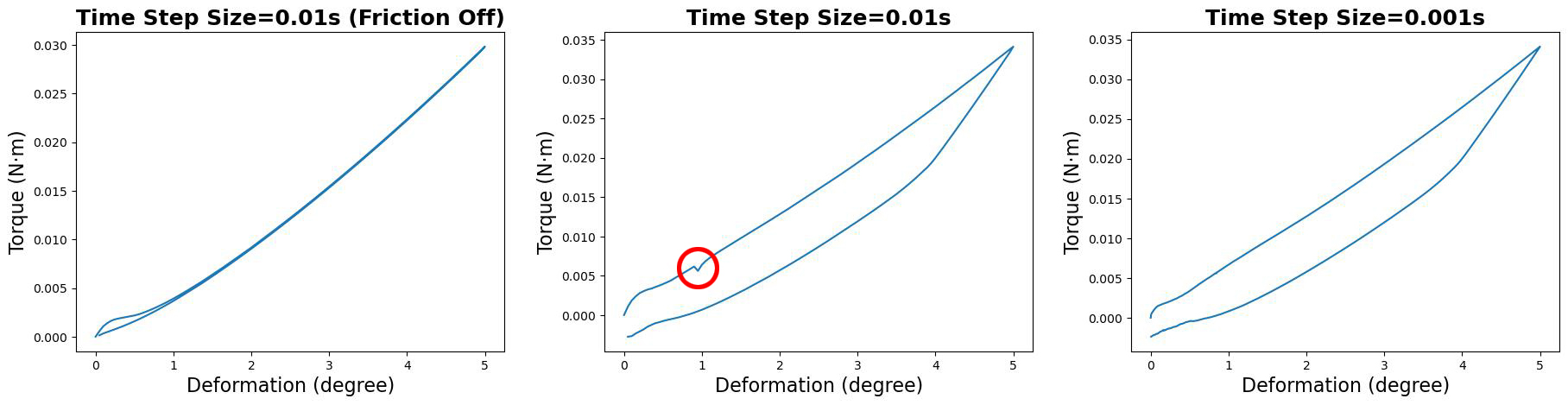}
    \caption{Our friction model may suffer from stability issues when using large time step size. The figures record the reaction torque when twisting cylindrical cloth samples as shown in~\Cref{fig:diff_material_twist}. The left figure shows the reaction torque when friction models are off. The middle figure shows an abrupt change appears after turning friction models on. This abrupt change disappears after reducing the time step size from 0.01s to 0.001s (shown in the right).}
    \label{fig:stability}
\end{figure*}

\begin{figure}[htb]
    \centering
    \includegraphics[width=0.48\textwidth]{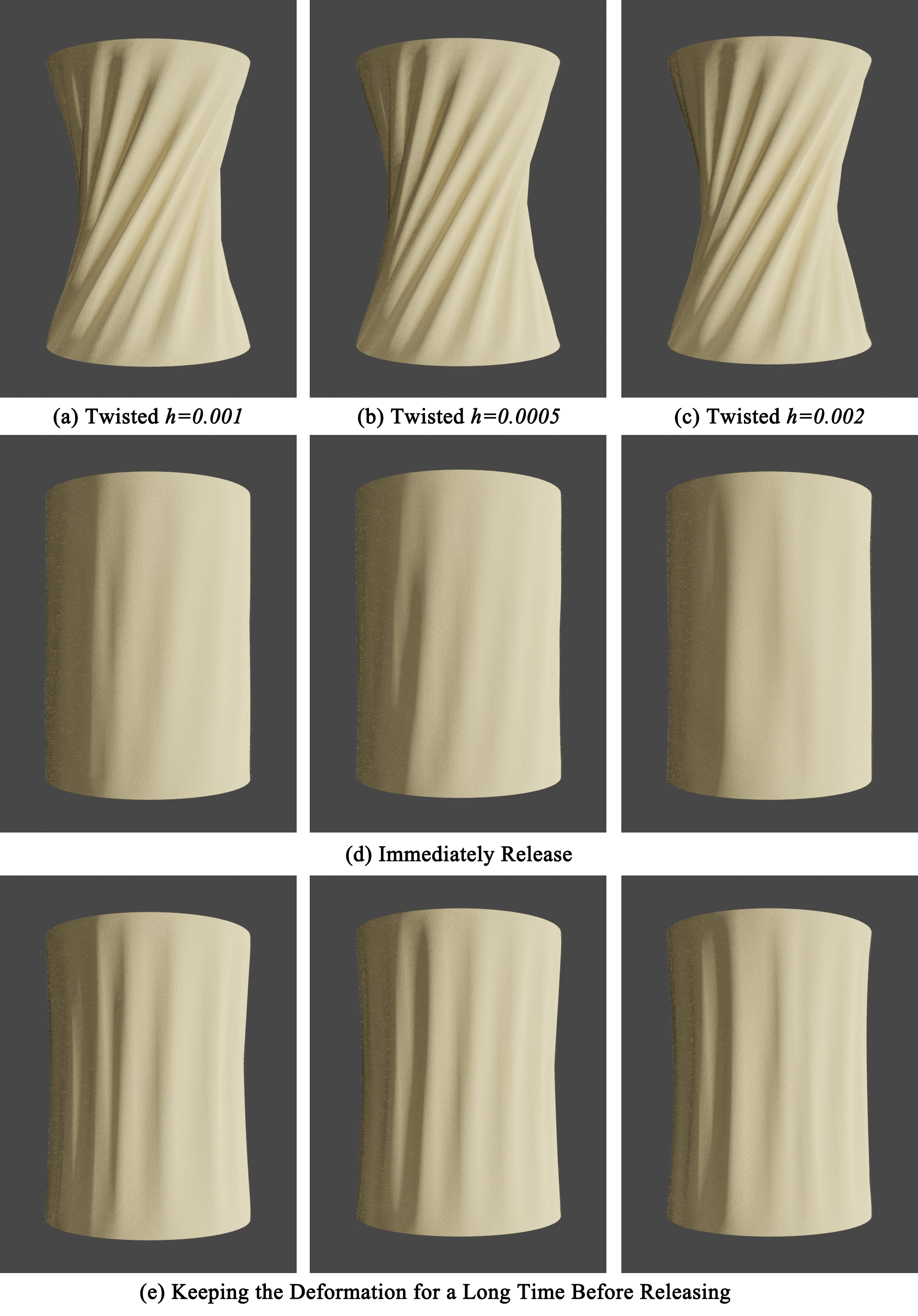}
    \caption{Our cloth wrinkle simulation method is time step size (i.e., $h$) independent.}
    \label{fig:ablation_step_size}
\end{figure}

We also show the same experiment when only plasticity is considered. A large initial deformation (\Cref{fig:ablation_friction_plastic} b-1) is followed by an immediate release (\Cref{fig:ablation_friction_plastic} b-2 and b-3) or a delayed release (\Cref{fig:ablation_friction_plastic} b-4 and b-5). Similar to the friction wrinkle, the wrinkle under the immediate release tends to be more reversible. Moreover, when comparing the friction wrinkle and the plastic wrinkle, it is easy to see that the plastic wrinkle is more persistent in all situations.  This experiment demonstrates that our model can successfully generate time-dependent friction and plastic wrinkles, and the high-fidelity is also shown in the varying reversibility of the wrinkles according to their causes and duration.

\textbf{Time Step Size} We twist three cloth samples of the same materials with different time step sizes to test if our cloth persistent wrinkle simulation method depends on time step size. The results are shown in~\Cref{fig:ablation_step_size}, which demonstrate the simulated time dependent wrinkles do not vary much with different $h$ values. 

\subsection{Stability Analysis}

Our anchor-based friction model might have stability issues when the time step size is large. To demonstrate this, we adopt the experiment scenario in Section 1.2, but twist a cloth sample by only 5 degrees. We only turn on the bending and stretching friction models to exclude the influences from plasticity. \Cref{fig:stability} records the reaction torque from the cloth samples when setting time step size to 0.01s and 0.001s respectively. The red circle in~\Cref{fig:stability} (middle) highlights an abrupt change which, however, does not appear when using pure elastic model~\Cref{fig:stability} (left) and disappears after reducing the time step size to 0.001s~\Cref{fig:stability} (right). Thus, we recommend reducing the time step size when our friction model encounters stability issues.

\section{Performance} 
We use the twisting experiments to show the performance of our simulator. This is to mainly measure the time needed for the internal friction and the plasticity, in the absence of other factors such as collisions and external frictions. \Cref{tab:time_profile} shows the time per simulation step when the cloth mesh has different resolutions: approximately 2K vertices, 7K vertices, and 27K vertices. Note our implementation is not optimized particularly for speed, hardware or parallel computing. The result demonstrates our friction and plastic models only cause a trivial addition to the computing and therefore they can be incorporated easily in simulation applications. 

\begin{table}[h]
    \centering
    \caption{Consumed time per simulation step (seconds/step) when using different resolution mesh. Turning on our friction and plastic model only trivially slow down the simulation.}
    \begin{tabular}{cccc}
        \toprule
         Model &  2K Mesh & 7K Mesh & 27K Mesh\\
         \midrule
         Friction/Plastic on & 0.279 & 1.319 & 12.064 \\
         Friction/Plastic off & 0.256 & 1.286 & 11.518 \\
         \bottomrule
    \end{tabular}
    \label{tab:time_profile}
\end{table}
\section{Cloth Physical Parameters}

The cloth physical parameters used in our experiments are shown in~\Cref{tab:sim_params} and~\Cref{tab:sim_params_garment}. In the tables, the parameters from $\rho$ to $K_{33}$ are the same as those in the pure elastic cloth simulators~\cite{volino2009simple, grinspun2003discrete}. The rest eight parameters are introduced by our friction and plastic models. Among them, $\tau_{f}$ and $\tau_{p}$ are used to control the speed of friction dwell and plastic hardening. For example, we set them to 30 seconds to make the friction dwell and hardening effect become obvious after 30 seconds. $k_{Friction}$ affects the strength of the friction force. To make the friction wrinkles obvious, $k_{Friction}$ should be large enough to produce sufficient friction force that can balance the elastic force. $\varepsilon_0$ and $\varepsilon_{inf}$ determines the maximum stick frictional strain and its maximum that can reach by frictional dwelling. To simulate obvious time dependence derived from friction, the difference between $\varepsilon_0$ and $\varepsilon_{inf}$ should be great. In our plastic model, $\varepsilon_{Y0}$ is the yield strain from which plastic deformation start occurring. Usually, $\varepsilon_{Y0}$ should be greater than $\varepsilon_0$ because cloth plastic deformations should be more difficult to occur than the cloth internal slip frictions. The hardening parameter $K_h$ decide how much deformation that surpasses the yield strain will flow to the plastic deformation, i.e., $\frac{K_b}{K_b + K_h}$. Therefore, if $K_h = K_b$, half of the deformation over the yield strain will become plastic deformation. Due to the time-dependent plastic hardening, $K_h$ can vary in the range defined by the parameters $K_{h0}$ and $g$: $[(1-g)K_{h0}, K_{h0}]$. Therefore, to make the time dependence caused by plasticity obvious, $g$ should be close to 1 and $K_{h0}$ should be similar to the elastic parameters.

\begin{table*}[htb]
    \centering
    \caption{Simulation parameters of simulate cotton (plain woven), denim (cotton twill), and polyester cloths.}
    \label{tab:sim_params}
    \begin{tabular}{cccc cccc cccc ccc}
        \toprule
        Cloth  &  $\rho$ & $3 \times K_b$ & $K_{11}$ 
        & $K_{22}$ & $K_{12}$ & $K_{33}$ & $3 \times K_{Friction} $ &$\varepsilon_0$ & $\varepsilon_{inf}$ 
         & $\tau_f$ & $K_{h0}$ & g 
        & $\tau_p$ & $\varepsilon_{Y0}$ \\
        \midrule
        Cotton &  0.06 & 5e-6  & 50.0 
        & 50.0 & 0.2 & 30.0  & 1e-5 & 0.1 & 1.7 
        & 30.0 & 5e-6 & 0.99 & 30.0 & 1.8 \\
        Denim &  0.25 & 1.2e-4  & 100.0 
        & 100.0 & 0.2 & 20.0 & 5e-5 & 0.1 & 1.8 
        & 30.0 & 1.2e-4 & 0.99 & 30.0 & 2.0 \\
        Polyester &  0.18 & 1.2e-4  & 50.0 
        & 50.0 & 0.2 & 30.0 & 1e-7 & 0.01 & 0.1 
        & 30.0 & 1.2e-4 & 0.99 & 30.0 & 3.0 \\
        \bottomrule
    \end{tabular}
\end{table*}

\begin{table*}[h]
    \centering
    \caption{Simulation parameters of simulate cotton, denim, and polyester trousers.}
    \label{tab:sim_params_garment}
    \begin{tabular}{cccc cccc cccc ccc}
        \toprule
        Cloth  &  $\rho$ & $3 \times K_b$ & $K_{11}$ 
        & $K_{22}$ & $K_{12}$ & $K_{33}$ & $3 \times K_{Friction} $ &$\varepsilon_0$ & $\varepsilon_{inf}$ 
         & $\tau_f$ & $K_{h0}$ & g 
        & $\tau_p$ & $\varepsilon_{Y0}$ \\
        \midrule
        Cotton &  0.1 & 1e-6  & 200.0 
        & 200.0 & 0.2 & 20.0  & 4e-6 & 0.1 & 1.2 
        & 30.0 & 1e-6 & 0.99 & 30.0 & 1.5 \\
        Denim &  0.2 & 3e-5  & 200.0 
        & 200.0 & 0.2 & 150.0 & 6e-5 & 0.2 & 1.2 
        & 30.0 & 3e-5 & 0.99 & 30.0 & 1.2 \\
        Polyester &  0.15 & 1e-6  & 100.0 
        & 100.0 & 0.2 & 20.0 & 7e-7 & 0.1 & 0.1 
        & 30.0 & 1e-6 & 0.99 & 30.0 & 3.1 \\
        \bottomrule
    \end{tabular}
\end{table*}

\section{Simulation Basics and Other Forces}

The cloth is discretized into a triangle mesh, whose state is represented by the positions and velocities of mesh vertices. Specially, given a cloth mesh with $v$ vertices, we denote its state at time $t$ by $\mathcal{S}_t = \{\mathbf{x}_t, \dot{\mathbf{x}}_t\}$, where $\mathbf{x} \in \mathrm{R}^{3v}$ and $\dot{\mathbf{x}} \in \mathrm{R}^{3v}$ denote the nodal position and velocity vector respectively. Given the initial state $\mathcal{S}_0$,  the cloth motion is governed by Newton's second law, $\mathbf{f} = \mathbf{M}\ddot{\mathbf{x}}$, where $\mathbf{M} \in \mathbf{R}^{3v \times 3v}$ is the lumped mass matrix~\cite{logan2022first} and $\mathbf{f} \in \mathrm{R}^{3v}$ is the net force vector: the combined force of the internal and the external forces at vertices. To solve $\mathbf{F} = \mathbf{M}\ddot{\mathbf{x}}$, we employ an implicit Euler formulation for stability under large time steps~\cite{baraff1998large}:
\begin{gather}
    \Delta \mathbf{x}_t = h \dot{\mathbf{x}}_{t} = h (\dot{\mathbf{x}}_{t-1} + \Delta \dot{\mathbf{x}}_t) \label{eq:var_pos} \\
    \Delta \dot{\mathbf{x}}_t = h \mathbf{M}^{-1} \mathbf{f}_t \label{eq:var_vel}
\end{gather}
where $h$ is the time step. Given the current state $\mathcal{S}_{t-1} = \{\mathbf{x}_{t-1}, \dot{\mathbf{x}}_{t-1}\}$, we can compute its future state $\mathcal{S}_{t} = \{\mathbf{x}_{t}, \dot{\mathbf{x}}_{t}\} = \{\mathbf{x}_{t-1} + \Delta \mathbf{x}_t, \dot{\mathbf{x}}_{t} + \Delta \dot{\mathbf{x}}_{t} \}$.
The key is to compute the force $\mathbf{F}_t = f (\mathcal{S}_t)$ where $f$ is a general force function of the cloth state. Although $\mathbf{F}_{t}$ is unknown without knowing the future state $\mathcal{S}_{t}$, it can be approximated by Taylor expansion:
\begin{equation}
    \label{eq:taylor}
    \mathbf{f}_{t} = \mathbf{f}_{t-1} 
    + \frac{\partial \mathbf{f}}{\partial \mathbf{x}} \Delta \mathbf{x}
    + \frac{\partial \mathbf{f}}{\partial \dot{\mathbf{x}}} \Delta \dot{\mathbf{x}}
\end{equation}
Combining Equation \ref{eq:var_pos}-\ref{eq:taylor}, we have the governing equation:
\begin{equation}
\label{eq:governingEq}
    \left(
    \mathbf{M} - h^2 \frac{\partial \mathbf{f}}{\partial \mathbf{x}}  
    - h \frac{\partial \mathbf{f}}{\partial \dot{\mathbf{x}}} 
    \right)
    \Delta \dot{\mathbf{x}}_{t}
    = h \left(
    \mathbf{f}_{t-1} + h \frac{\partial \mathbf{f}}{\partial \mathbf{x}}\dot{\mathbf{x}}_{t-1}
    \right)
\end{equation}
which can be solved by an iterative solver, e.g., Conjugate Gradient \cite{shewchuk1994introduction}.


In our cloth simulation, the in-plane stretching and out-of-plane bending forces tend to keep the cloth in its rest state. For bending, we use a linear elastic model to define the bending energy density: $\psi_b(\varepsilon_b)=\frac{1}{2}K_b \varepsilon_b^2$ where $K_b$ is the bending stiffness and $\varepsilon_b$ is the bending strain. Therefore, the elastic bending stress is $\sigma_b = \frac{\text{d} \psi_b(\varepsilon_b)}{\text{d} \varepsilon_b} = K_b \varepsilon_b$. The bending energy of cloth is the integral of the bending energy density over the cloth's surface, i.e., $\mathcal{W}_b=\int \psi(\varepsilon_b) \text{d}A$ where $A$ is the area of the cloth. On a discretized mesh, the integral of overall bending energy is approximated by the sum of the bending energies around all edges: $\mathcal{W}_b=\sum W_b$. We adopt the method in \cite{grinspun2003discrete} which uses the mean-curvature to define the bending strain: $\varepsilon_b = 3 \frac{\theta - \bar{\theta}}{\bar{\mathrm{H}}}$ where $\theta$ is the Dihedral angle between two adjacent triangles and $\bar{\theta}$ is the rest Dihedral angle. $\bar{\mathrm{H}}$ is the average height of the two triangles. The bending energy around a edge is
\begin{equation}
    W_b = A_b \int^{\varepsilon_b}_{0} \sigma_b \text{d} \varepsilon_b = A_b \int^{\varepsilon_b}_{0} K_b \varepsilon_b \text{d} \varepsilon_b = \frac{1}{2} A_b \sigma_b \varepsilon_b
\end{equation}
The nodal bending force is defined as the derivatives of the energy
\begin{equation}
    \label{eq:bending_force}
    \mathbf{f}_b = - \frac{\partial W_b}{\partial \mathbf{x}_b} = - A_b \sigma_b \frac{\partial \varepsilon_b}{\partial \mathbf{x}_b} = - 3 \frac{A_b \sigma_b}{\bar{\mathrm{H}}} \frac{\partial \theta}{\partial \mathbf{x}_b}
\end{equation}
where $W_b$ is the bending energy, $\varepsilon_b^{curr}$ denotes current bending strain, and $A_b = \frac{1}{3}l\bar{\mathrm{H}}$ where $l$ is the rest length of the common edge shared by the two triangles. $\mathbf{x}_b \in \mathcal{R}^{12}$ is the vertex position vector of the two triangles.

For stretching, we adopt hyperelastic Saint-Venant-Kirchhoff (StVK) constitutive model~\cite{volino2009simple}, which encodes the material nonlinearity in the strain. It measures cloth stretching deformation by the Green-Lagrange strain tensor $\mathbf{E}$:
\begin{equation}
    \mathbf{E} = \frac{1}{2} (\mathbf{F^\top F} - \mathbf{I}) =
    \begin{bmatrix}
        \varepsilon_{uu}, \varepsilon_{uv}\\
        \varepsilon_{uv}, \varepsilon_{vv}
    \end{bmatrix}
\end{equation}
where $\mathbf{F}^{3\times2}$ denote the cloth's deformation gradient and $\mathbf{I}^{2\times2}$ is an identity matrix. In Voigt Notation, it can be denoted by the vector $\boldsymbol{\varepsilon}_s = (\varepsilon_{uu}, \varepsilon_{vv}, \varepsilon_{uv})$. Cloths, e.g., woven fabrics, usually exhibit distinctive stretching mechanical properties in the warp and the weft direction, and they are usually modeled as orthotropic materials~\cite{boisse2001analyses}. Therefore,  $\varepsilon_{uu}$, $\varepsilon_{vv}$, and $\varepsilon_{uv}$ represent the tensile strains along the warp, the weft, and the diagonal direction (shearing strain), respectively. The stretching constitutive equation is:
\begin{equation}
    \boldsymbol{\sigma}_s = 
    \begin{bmatrix}
        \sigma_{uu} \\ \sigma_{vv} \\ \sigma_{uv}
    \end{bmatrix}
    = \begin{bmatrix}
        k_{11} & k_{12} & 0 \\
        k_{12} & k_{22} & 0 \\
        0 & 0 & k_{33} \\
    \end{bmatrix}
    \begin{bmatrix}
        \varepsilon_{uu} \\ \varepsilon_{vv} \\ \varepsilon_{uv}
    \end{bmatrix}
    = \mathbf{K}_s \boldsymbol{\varepsilon}_s
\end{equation}
where $\mathbf{K}_s$ is the stretching stiffness matrix in which $k_{11}$, $k_{22}$,  $k_{33}$, and $k_{12}$ are the warp/weft/shear stretching stiffness and Poisson's ratio~\cite{wang2011data}. The energy density function of the StVK model is defined as
\begin{align}
    \psi_s(\mathbf{E}) &= \frac{1}{2}(\lambda + 2\mu)\rom{1}_E^2 -2\mu \rom{2}_E\\
    &=\frac{\lambda + 2\mu}{2} \varepsilon_{uu}^2 + \frac{\lambda + 2\mu}{2} \varepsilon_{vv}^2 + \lambda \varepsilon_{uu}\varepsilon_{vv} - 2\mu \varepsilon_{uv}^2
\end{align}
where $\lambda$ and $\mu$ are lam\'e constants which can be converted to the entries in $\mathbf{K}_s$: $k_{11}=k_{22}=\lambda + 2\mu$, $k_{12}=\lambda$, and $k_{33}=4\mu$. $\rom{1}_E=\text{Tr}(E)=\varepsilon_{uu} + \varepsilon_{vv}$ and $\rom{2}_E=\frac{1}{2}(\text{Tr}(E)^2-\text{Tr}(E^2))=\varepsilon_{uu}\varepsilon_{vv}-\varepsilon_{uv}^2$ are the principle invariants of the Green-Lagrange strain tensor. The stretching energy of the cloth is $\mathcal{W}_s=\int \psi_s(E)\text{d}A$ which can approximated by sum of all triangle faces stretching energy: $\sum W_s$ in a discretized mesh. The nodal stretching force is the partial derivative of the stretching energy over a triangular face w.r.t. the vertex position:
\begin{gather}
    W_s = \int \psi_s(\mathbf{E}) \mbox{d} A_s
    = A_s \psi_s(\mathbf{E}) \\
    \label{eq:stretching_force}
    \mathbf{f}_s = - \frac{\partial W_s}{\partial \mathbf{x}_s} 
    = - A_s  \boldsymbol{\sigma}_s \frac{\partial \boldsymbol{\varepsilon}_s}{\partial \mathbf{x}_s}
\end{gather}
where $A_s$ is the rest area of a triangle and $\mathbf{x}_s \in \mathcal{R}^{9}$ is the position vector of the triangle's vertices. Please refer to work proposed by~\cite{volino2009simple} for $\frac{\partial \boldsymbol{\varepsilon}_s}{\partial \mathbf{x}_s}$ and the force Jacobians.

Other than the internal forces, we also consider external forces which include gravity, collision, external friction, and handle force. Gravity is a body force and is applied to all lump masses. In addition, we adopt~\cite{bridson2002robust} to handle self-collisions and external collisions. The method uses repulsion forces to separate the elements in proximity. Moreover, if there is relative sliding between elements, external friction forces are introduced to prohibit relative motions. Finally, the handle force is intended for controlling the cloth and derives from a penalty-based method to pin vertices at specified locations:
\begin{equation}
    \mathbf{f}_h = k_h (\mathbf{x} - \mathbf{x}_h)
\end{equation}
where $k_h$ is the handle stiffness and $\mathbf{x}_h$ specifies the anchor positions in space.

\section{Force Derivatives}

\textbf{Internal Friction} Similarly to bending force:
\begin{equation}
    W_b = A_b \int^{\varepsilon_b}_{0} \sigma_b \text{d} \varepsilon_b = A_b \int^{\varepsilon_b}_{0} K_b \varepsilon_b \text{d} \varepsilon_b = \frac{1}{2} A_b \sigma_b \varepsilon_b
\end{equation}
\begin{equation}
    \mathbf{f}_b = - \frac{\partial W_b}{\partial \mathbf{x}_b} = - \frac{1}{2} A_b \sigma_b \frac{\partial \varepsilon_b}{\partial \mathbf{x}_b} = - 3 \frac{A_b \sigma_b}{\bar{\mathrm{H}}} \frac{\partial \theta}{\partial \mathbf{x}_b} \mbox{,}
\end{equation}
we first define a friction energy on the bending edge:
\begin{align}
    W_{friction} &= A_b\int_{\bar{\varepsilon}}^{\varepsilon}\sigma_{friction} d \varepsilon_b = A_b \int_{\bar{\varepsilon}}^{\varepsilon} K_{friction} \Delta \varepsilon_b d \varepsilon_b \nonumber \\
    &= A_b (\frac{1}{2} K_{friction} \varepsilon_b^2 
    - K_{friction} \bar{\varepsilon}_b \varepsilon_b) 
\end{align}
Therefore, the corresponding friction forces on the four vertices around the bending edge are:
\begin{align}
    \label{eq:friction_force}
    \mathbf{f}_{friction} &= -\frac{\partial W_{friction}}{\partial \mathbf{x}} \notag \\
    &= -\frac{\partial W_{friction}}{\partial \varepsilon_b} \frac{\partial \varepsilon_b}{\partial \mathbf{x}} \notag \\
    &= -A_b K_{friction} (\varepsilon_b - \bar{\varepsilon}_b)
    \frac{\partial \varepsilon_b}{\partial \mathbf{x}}
\end{align}
To solve \Cref{eq:governingEq}, we need the force Jacobian w.r.t the vertex positions:
\begin{equation}
    \label{eq:force_pos_Jaco}
    \frac{\partial \mathbf{f}_{friction}}{\partial \mathbf{x}}
    = -A_b K_{friction} 
    \left(
        \frac{\partial^2 \varepsilon_b}{\partial \mathbf{x}^2}  +
        \frac{\partial \varepsilon_b}{\partial \mathbf{x}} 
        {\frac{\partial \varepsilon_b}{\partial \mathbf{x}}}^{\top}
    \right)
\end{equation}
By adopting Gauss-Newton approximation, we set the bending strain's second-order derivative to zero, i.e., $\frac{\partial^2 \varepsilon_b}{\partial \mathbf{x}^2}=0$. On the one hand, our simulator can avoid the indefinite $\theta$'s Hessian which would make the left-hand side matrix in \Cref{eq:governingEq} non-positive definite~\cite{tamstorf2013discrete}, on the other hand, can reduce the computation consumption.

\textbf{Plastic Model} After integrating the plastic model, the bending nodal force defined in the main paper
\begin{equation}
    \mathbf{f}_b = - \frac{\partial W_b}{\partial \mathbf{x}_b} = - \frac{1}{2} A_b \sigma_b \frac{\partial \varepsilon_b}{\partial \mathbf{x}_b}
\end{equation}
becomes
\begin{equation}
    \mathbf{f}_b = 
    -\frac{1}{2} A_b \sigma_b \frac{\partial \varepsilon_b}{\partial \mathbf{x}} = 
    -\frac{1}{2} A_b K_b (\varepsilon_b - \varepsilon_p) \frac{\partial (\varepsilon_b - \varepsilon_p)}{\partial \mathbf{x}}
\end{equation}
In each step, the variation of $\varepsilon_p$ consists of two parts: immediate plastic flow and time-dependent hardening. The former updates immediately and then is taken as a constant new rest strain, so its derivative is zero. The second part is modeled by \Cref{eq:Kh} which is not a function of x. Therefore, $\frac{\partial \varepsilon_p}{\partial \mathbf{x}}$ is zero. Thus, the bending nodal force is
\begin{equation}
    \mathbf{f}_b = 
    -\frac{1}{2} A_b K_b (\varepsilon_b - \varepsilon_p) \frac{\partial \varepsilon_b}{\partial \mathbf{x}}
\end{equation}
where we avoid computing complex derivatives and do not observe any negative impact on the simulation stability in practice.

\section{Algorithms}

The algorithms of our time-dependent friction and plastic model are shown in \Cref{alg:friction} and \Cref{alg:plastic} respectively. 

\begin{algorithm}[htb]

    \caption{Time-dependent Friction} 
    \label{alg:friction}

    \SetKwProg{Fn}{Function}{}{end}
    \SetKwFunction{MyFunction}{Friction}
    
    \Fn{\MyFunction($\varepsilon_{n},\varepsilon_{n-1}, \bar{\varepsilon}$)}
    {
        $\dot{\varepsilon} \gets \frac{\varepsilon_{n} - \varepsilon_{n-1}}{h}$\;
        $\varepsilon_{thres} = \varepsilon_{inf} - (\varepsilon_{inf} - \varepsilon_{0}) e^{-t_{stick} / \tau_f}$\;
        $\Delta \varepsilon  = \varepsilon_{n} - \bar{\varepsilon}$\;
        \If(\tcc*[f]{Slip Friction}){$|\Delta \varepsilon| > \varepsilon_{thres}$}{
            $\bar{\varepsilon} \leftarrow \bar{\varepsilon} + sign(\Delta \varepsilon ) * (| \Delta \varepsilon| - \varepsilon_{thres} )$ \tcc*{Equation 4}
        }
        \textbf{compute} $\mathbf{f}, \frac{\partial \mathbf{f}}{\partial \mathbf{x}}$ \tcc*{Equation 23 and 24}
        \textbf{return} $\mathbf{f}, \frac{\partial \mathbf{f}}{\partial \mathbf{x}}$\;
    }       
\end{algorithm}

\begin{algorithm}[htb]
    \caption{Time-dependent Hardening Plastic Model}
    \label{alg:plastic}

    \SetKwProg{Fn}{Function}{}{end}
    \SetKwFunction{MyFunction}{PlasticBending}

    \Fn{\MyFunction($\varepsilon_{n}, \varepsilon_p, \varepsilon_Y$)}{
        $\varepsilon_e = \varepsilon_{n} - \varepsilon_p$ \;
        \If(\tcc*{Plastic Deformation}){$ |\varepsilon_e| > \varepsilon_{Y} $}{
            \eIf{$sign(\varepsilon_e) = sign(\varepsilon_p)$}{
                $t_{plastic} \gets t_{plastic} + h $\;
            }{
                $t_{plastic} \gets 0 $\;
            }
            $K_h = K_{h0} (1 - g(1-e^{(-t_{plastic}/ \tau_p)}))$ \tcc*{Equation 7}
            $\varepsilon_{hp} \gets \varepsilon_{hp} + \frac{K_b}{K_b + K_h}(|\varepsilon_e| - \varepsilon_Y)$ \tcc*{Equation 8}
            $\varepsilon_{p} \gets \varepsilon_{p} + sign(\varepsilon_e)\frac{K_b}{K_b + K_h}(|\varepsilon_e| - \varepsilon_Y)$ \tcc*{Equation 9}
            $\varepsilon_Y = \varepsilon_{Y0} + \varepsilon_{hp} \frac{K_h}{K_b}$ \tcc*{Equation 10}
            $\varepsilon_e \gets \varepsilon_Y $\;
        }
        \textbf{Compute} $\mathbf{f}, \frac{\partial \mathbf{f}}{\partial \mathbf{x}}$\;
        \tcc{Equation 3 and \cite{grinspun2003discrete}}
        \textbf{return }$\mathbf{f}, \frac{\partial \mathbf{f}}{\partial \mathbf{x}}$\;
    }  
\end{algorithm}

\section{Implementation}

Our implementation is in MSVC(10.0.19041.0) C++ with OpenMP for CPU parallel computing. We use the Eigen library \cite{eigenweb} for matrix calculation and solving the governing equation. Our experiments are conducted on a PC with an Intel i7-12700H 2.3GHz CPU and 16GB 4800MHz RAM. 

For simplicity, we use the Dihedral angle $\theta$ to define the slide friction threshold $\varepsilon_{thres}$ and the yield strain $\varepsilon_Y$, rather than the bending strain $\varepsilon_b$. As we use fixed and uniform discretization to triangulate cloth meshes in our experiments, adopting either of these two metrics does not affect the physical system. The bending strain, i.e., $\varepsilon_b=3 \frac{\theta - \bar{\theta}}{e}$, is just the scaled Dihedral angle. $\varepsilon_b$ would become very large when using the fine-resolution meshes in which $e$ is very small. In comparison, the Dihedral angle is more straightforward to understand. When adopting non-uniform discretization or adaptive remeshing, it would be more feasible to use curvature-based metrics, e.g., bending strain $\varepsilon_b$, to define the thresholds because they are agnostic to the meshes' topology.

The parameters $\tau_f$ and $\tau_p$ allow us to define the variation rate of $\varepsilon_{thres}$ and $K_h$ with respect to the deformation duration. Remember $\varepsilon_{thres}$ and $K_h$ control the friction dwell effect and plastic hardening effect respectively which further decide how fast the friction/plastic wrinkles appear in time. To show the time-dependence, theoretically our simulator can run for a long time to simulate e.g. slow hardening processes. However, this might not be ideal in applications. Therefore, we tune the parameters given a fixed deformation duration (i.e. how long a deformation is kept) for different materials. In particular, given the same deformation duration, the smaller the $\tau$'s are, the more obvious the wrinkles tend to be. Therefore, we can flexibly determine a deformation duration that is long enough for the friction/plastic wrinkles to develop, as long as that duration is much greater than the $\tau$'s. In our experiments, we set the $\tau$'s to 30 seconds and the long deformation duration to 500 seconds which is sufficient to observe the wrinkle time dependence. In addition, we can increase $t_{stick}$ and $t_{plastic}$ by larger time steps to avoid running the simulation for too long. For instance, in our experiments, we increase $t_{stick}$ and $t_{plastic}$ by 10 seconds in every simulation step when keeping the cloth's deformation. This enables us to only run 50 steps before $t_{stick}$ and $t_{plastic}$ reach 500 seconds.

\end{appendices}


\bibliographystyle{eg-alpha-doi} 
\bibliography{egbibsample}       


\end{document}